\documentclass[12pt,preprint]{aastex} 

\begin{document} 

\title 
{The $K$--selected Butcher-Oemler Effect} 

\author 
{Roberto De Propris} 

\affil 
{Research School of Astronomy \& Astrophysics, Australian National 
University, Weston Creek, ACT, 2611, Australia} 
\email{propris@mso.anu.edu.au }

\author 
{S. A. Stanford\altaffilmark{1,2}} 
\affil 
{Physics Department, University of California at Davis, Davis, CA 95616} 
\email{adam@igpp.ucllnl.org} 
\altaffiltext{1} 
{Institute of Geophysics and Planetary Physics,  Lawrence Livermore National 
Laboratories,  Livermore, CA, 94550} 
 
\author 
{Peter R. Eisenhardt\altaffilmark{2}} 
\affil 
{MS 169-327, Jet Propulsion Laboratory, 
California Institute of Technology, 
4800 Oak Grove Drive, Pasadena, CA, 91109} 
\email{prme@kromos.jpl.nasa.gov} 
 
\author 
{Mark Dickinson\altaffilmark{2}} 
\affil 
{Space Telescope Science Institute, 3700 San Martin Drive, Baltimore, MD, 21218} 
\email{med@stsci.edu}

\altaffiltext{2}{Visiting Astronomer, Kitt Peak National Observatory, National 
Optical Astronomy Observatories, which is operated by the Association of 
Universities for Research in Astronomy, Inc. (AURA) under cooperative agreement 
with the National Science Foundation.}

\begin{abstract} 
 
We investigate the Butcher-Oemler effect using samples of galaxies
brighter than observed frame $K^*+1.5$ in 33 clusters at $0.1 \lesssim
z \lesssim 0.9$.  We attempt to duplicate as closely as possible the
methodology of Butcher \& Oemler.  Apart from selecting in the
$K$-band, the most important difference is that we use a brightness
limit fixed at 1.5 magnitudes below an observed frame $K^\ast$ rather
than the nominal limit of rest frame $M(V) = -20$ used by Butcher \&
Oemler. For an early type galaxy at $z=0.1$ our sample cutoff is 0.2
magnitudes brighter than rest frame $M(V)=-20$, while at $z=0.9$ our
cutoff is 0.9 magnitudes brighter.  If the blue galaxies tend to be
faint, then the difference in magnitude limits should result in our
measuring lower blue fractions.  A more minor difference from the
Butcher \& Oemler methodology is that the area covered by our galaxy
samples has a radius of 0.5 or 0.7 Mpc at all redshifts rather than
$R_{30}$, the radius containing 30\% of the cluster population. In
practice our field sizes are generally similar to those used by
Butcher \& Oemler.

We find the fraction of blue galaxies in our $K$-selected samples to
be lower on average than that derived from several optically selected
samples, and that it shows little trend with redshift.  However, at
the redshifts $z < 0.6$ where our sample overlaps with that of Butcher
\& Oemler, the difference in $f_B$ as determined from our $K$-selected
samples and those of Butcher \& Oemler is much reduced. The large
scatter in the measured $f_B$, even in small redshift ranges, in
our study indicates that determining the $f_B$ for a much larger 
sample of clusters from $K$-selected galaxy samples is important.

As a test of our methods, our data allow us to construct
optically-selected samples down to rest frame $M(V) =-20$, as used by
Butcher \& Oemler, for four clusters that are common between our
sample and that of Butcher \& Oemler.  For these rest $V$ selected
samples, we find similar fractions of blue galaxies to Butcher \&
Oemler, while the $K$ selected samples for the same 4 clusters yield
blue fractions which are typically half as large.  This comparison
indicates that selecting in the $K$-band is the primary difference
between our study and previous optically-based studies of the Butcher
\& Oemler effect.

Selecting in the observed $K$-band is more nearly a process of selecting
galaxies by their mass than is the case for optically-selected samples.
Our results suggest that the Butcher-Oemler effect is at least partly due to
low mass galaxies whose optical luminosities are boosted.  These lower
mass galaxies could evolve into the rich dwarf population observed in 
nearby clusters.
 
\end{abstract} 
\keywords{galaxies:formation and evolution --- galaxies: clusters} 
 
\section{Introduction} 
 
In the late 1970's a variety of new imaging technologies were being 
tried out, with an order of magnitude or more better sensitivity than
photographic plates. Butcher \& Oemler's (1978) observations of two
galaxy clusters at $z \sim 0.4$ using the ISIT Vidicon on the KPNO 2.1~m
altered the landscape for studies of distant galaxies, providing the first
clear indication of dramatic changes in galaxy properties, and in the 
unexpectedly recent past. Butcher \& Oemler (1984; hereafter BO84) confirmed
the reality of the effect by addressing a number of concerns about 
systematic effects in the analysis procedure, and by broadening the sample
both in number and redshift.  The Butcher-Oemler (BO) effect---the discovery 
that galaxy clusters at $z > 0.2$ contain a higher fraction of blue 
galaxies ($f_B$) than do nearby galaxy clusters---inspired an 
empirical approach to galaxy evolution which continues to the present 
day. 
  
The origin of the BO effect is important to discern. Larson, Tinsley, 
\& Caldwell (1980) suggested  that S0s could be later type disks transformed 
by the cluster environment into earlier types.  The fraction of blue galaxies 
in $z \sim 0.4$ clusters is qualitatively similar to the excess of lenticulars 
(S0) in nearby clusters relative to those at $z \sim 0.5$ \citep{dre97}, 
supporting the idea that infalling field spirals are transformed into S0s. 
Morphological transformation from spiral to S0 may be induced by 
starbursts occurring upon cluster infall and leading to the rapid 
exhaustion of gas reservoirs \citep{dg83,bar96,pog99}, by truncation 
of normal star formation after infall \citep{abr96,mor98}, or by a 
gradual decline in star formation as the dark haloes surrounding 
spirals are removed and gas supplies can no longer be replenished by 
cooling and infall \citep{ltc80,bal99}. These mechanisms also provide 
natural explanations for the origin of the morphology-density relation 
\citep{dre80} and agree with the apparent preference of the blue 
galaxies for the outskirts of clusters (BO84). The starburst
explanation is supported by spectroscopy of the blue galaxies 
indicating that they are seen during or shortly after an episode of 
star formation \citep{cs87,co94,co98} (hereinafter CS87, C94, and C98 
respectively). Finally, $HST$ images indicate that the blue galaxies 
are predominantly normal late-type disks, with some tendency to be 
involved in interactions (Lavery \& Henry 1988, Oemler et al. 1997,
C94, C98). 

On the other hand, \cite{ros97} and C98 conclude that many of
the most actively star forming blue galaxies are actually low luminosity
systems, temporarily brightened by starbursts, which will fade from view and
evolve not into S0s but into dwarf galaxies. In this bursting dwarf
scenario the optical luminosities of such galaxies are boosted to
enter the luminosity cut defined by BO84.  This raises the possibility
that at least part of the BO effect is a result of photometric
selection.  In BO84 selection is carried out in optical passbands,
which could magnify the effect of even minor starbursts (even though
the BO84 passbands correspond approximately to the rest-frame $V$
band).

Our understanding of the physical mechanisms responsible for the BO effect
remains contentious.  Part of the problem is simply that there is a wide
range in measured blue fractions at a given redshift.  This could be 
explained by the larger problem that the cluster samples usually used 
in studies of the BO effect do not consist of the same kind of clusters 
over the redshift range of interest.  Andreon \& Ettori's (1999) analysis 
of the clusters in BO84 using X-ray data indicate that the latter's 
results are biased.  They found that the $L_x$ of the BO84 sample 
increases with redshift, whereas there is no evidence for evolution in 
the X-ray luminosity function up to $z \sim 0.8$.  When \cite{ae99} add 
X-ray selected clusters to the BO84 sample, the trend of the blue 
fraction with redshift is much reduced.  \cite{maca00} found similar 
results, underscoring the need for a study of the BO effect using a 
well-defined sample of clusters, over a large redshift range.  Further
illustrating the importance of sample selection, \cite{margo01} found
a strong correlation between cluster richness and the blue fraction, in
the sense that $f_B$ is higher for poorer clusters, using a large
sample of 295 Abell clusters. 

Selection of galaxy samples in the near-infrared as opposed to the 
optical should result in samples more representative by stellar {\it mass} 
\citep{aes91,gpb96,sed98}. \cite{sed98} and \cite{dep99} have shown that
the luminosities and colors of galaxies selected in the $K$ band are not
strongly dependent on their environment. Therefore $K$-selected samples 
can be used to constrain the masses of the blue galaxies causing the BO 
effect and investigate if these objects evolve into S0's or dwarfs. 

We present here a study of the BO effect in distant clusters using 
galaxies selected in the $K$ band. Our sample, observations and data
reduction are described in the next section. The analysis and the main
results are presented in Section~\ref{secana}. We discuss our findings
in Section~\ref{secdisc}. For consistency with BO84, we adopt a 
cosmology with $H_0 = 50$ km s$^{-1}$ Mpc$^{-1}$ and $q_0=0.1$. 

\section{Data} \label{secobs} 

Our sample is the heterogeneous set of clusters which were studied for luminosity 
function evolution by \cite{dep99}.  For this sample, we have $JK$ and 
two optical bands of imaging which reach at least to 1.5 magnitudes below an 
evolving $K^\ast$ at the 5$\sigma$ level.  The optical bands change with 
redshift, bracketing the 4000 ${\rm \AA}$ break as closely as possible. 
Catalogs for these clusters were generated from the $K$-band images. 
The observations, data reduction, and photometry for this sample are 
presented in \cite{sta02}. 

\section{Analysis} \label{secana}

In our analysis we are attempting to duplicate the methodology of BO84 
as closely as possible.  To that end, we begin with a summary of the 
cluster sample and procedures used by BO84.  Their sample consisted of 
33 clusters spanning the range $0.003\ {\rm (Virgo)} < z < 0.54$ drawn 
mostly from the Abell catalog.  The data used by BO84 came from three main 
sources.  Except for CL0016+16 (photographic plates were obtained by 
Koo 1981), the most distant clusters were observed with the ISIT 
Vidicon camera on the Kitt Peak 2.1~m telescope in the $V$ and $R$ 
bands covering areas of $\sim$6.2 arcmin$^2$ down to a limit of $R 
\sim 22.0$.  The photometry for their intermediate redshift clusters 
was obtained using $J$ and $F$ plates on the 4~m telescopes at KPNO 
and CTIO and covered areas of 55 arcmin$^2$ down to a red magnitude of 
22.  The photometry for the low redshift clusters was obtained from 
$J$ and $F$ plates done at the Palomar 1.2~m Schmidt and covered areas 
of radius 1.5 Mpc. 

BO84 defined `blue' galaxies as (i) objects lying within a radius 
containing 30\% of the cluster population (R$_{30}$); (ii) brighter 
than a no-evolution $M_V=-20$; and (iii) bluer by 0.2 magnitudes in 
rest frame $B-V$ than the red sequence defined by the cluster E/S0 
galaxies.  BO84 derived the slope and intercept of the color-magnitude 
relation from their data or from the unevolved slope of the relation in 
nearby clusters \citep{sv78} and calculated their $B-V$ color offset 
using spectral energy distributions from \cite{cww80} and assuming no 
evolution. 

Because of the smaller size of infrared detectors, we are generally 
unable to cover enough field to trace the surface density distribution 
of galaxies and derive $R_{30}$ directly from our data. Therefore we 
adopt circular apertures of radius 0.5 and 0.7 Mpc centered on the 
brightest cluster galaxy.  The $R_{30}$ listed in Table 1 of BO84 are 
between $\sim$1 and $\sim$5 arcmin with an average of 2.2 arcmin for 
the clusters at $z > 0.2$ (see our Table~\ref{fbcomp} for the $R_{30}$ 
of the comparison clusters). A radius $= 0.7$ Mpc is very similar to 
this average $R_{30}$ in the redshift range where our sample overlaps 
with that of BO84.

As for the magnitude limit for our samples, first we need to convert
from the optical band used in BO84 to the $K$-band we are using for
galaxy selection. We calculate that $M_V=-20$ corresponds to
$M_K=-23.05$ for early-type galaxies in the present epoch, based on
data from the $UBVRIzJHK$ survey of the Coma cluster by \cite{eis02a}.
Our data on the distant clusters do not reach $M_K=-23.05$ in all
cases.  We choose to use a magnitude cut at $K^*+1.5$, where $K^*$ has
been determined from our data on each cluster \citep{dep99}.
Table~\ref{tab1} gives these $K^\ast$, along with both the limiting
$K$ actually used with our data in calculating blue fractions, the
observed frame $K$ magnitude that is equivalent to the nominal
$M(V)=-20.0$ for an elliptical galaxy assuming pure $k-$correction
(using the models of Poggianti 1997) of the observed colors of Coma
early-types, and the observed frame $K$ equivalent to the {\it actual}
limits in $M(V)$ used by BO84 (see Table~\ref{fbcomp}). For $0.1 < 
z < 0.3$ our limiting magnitudes are somewhat brighter than those used 
in BO84, a point we will return to later.  While different
from the fixed absolute magnitude cut used by BO84, our variable
magnitude limit accounts for the passive evolution seen in cluster
galaxies (e.g., Stanford et al. 1998).  The single absolute magnitude limit
used by BO84 translates into a cut relative to $M^\ast$ that changes
with redshift, thus possibly including variable proportions of giant
and dwarf galaxies into their samples, whereas our magnitude limit
is more likely to sample similar populations at all redshifts.

To determine which galaxies in our samples defined by the magnitude
and area limits described above are blue we follow the methodology of
BO84.  First we correct the galaxy colors for the color-magnitude
correlation of E/S0 galaxies by fitting the optical and optical-IR
colors with a robust linear least squares routine with brightest
cluster galaxies excluded from the fits. This algorithm minimizes
least absolute deviation (rather than its square) with iterative
rejection (Applied Statistics algorithm \# 132) and therefore reduces
the influence of outliers. We carry out this fit within the 0.5 Mpc
region to maximize the strength of the cluster red sequence.  We
inspect the fit by eye and, occasionally, intervene manually to
produce a more acceptable relation (i.e. one where the mode of the
marginalized color distribution is as close as possible to 0).  The
fit should be done only to the early-type galaxies but we do not have
morphological information on all the clusters in our sample.  The
color-magnitude diagrams and the best fits are shown in Figure 1 for
both the optical-IR and the pure optical colors.

In order to set a color boundary that defines the blue galaxies for
our chosen passbands, we need to transform the rest frame $(B-V)=0.2$
offset into a difference in the observed colors ($B-R$, $g-R$, $V-I$,
$R-I$, $R-K$ and $I-K$ as described below) at the redshift of each
cluster. We follow the procedure outlined by BO84 and \cite{sma98} and
use E, Sa and Sc spectral energy distributions from \cite{pog97}.  We
find that a mixture of 55\% Sa and 45\% Sc yields a spectrum with a
color difference $\Delta(B-V)=0.2$ from an E model.  We use this mixed
spectrum to calculate the corresponding $\Delta\ (color)$ (for our
observed colors) at $z=0$. We then use the $k-$ corrections given in
\cite{pog97} (for no-evolution models) to calculate a color difference
as a function of redshift (i.e., we compute the no-evolution colors
for the E model and for the Sa+Sc mixture). For example, the
difference corresponding to an offset of $(B-V)=0.2$ in the observed
$V-I$ at $z=0$ is 0.21 mag. For a solar metallicity elliptical in
e.g.\ GHO1601+4253 the $k-$correction in $V$ is 1.77 mag and 0.41 in
$I$ (Poggianti 1997); for the composite Sa+Sc spectrum the
$k-$correction is 1.24 in $V$ and 0.21 in $I$. Therefore the
$k-$corrected difference corresponding to $(B-V)=0.2$ is 0.54
magnitudes in $V-I$ at $z=0.54$.

We remove contamination from field galaxies by using the SPICES survey
\citep{spices2,eis02b}. SPICES is an imaging survey of four fields
(Cetus, Lynx, Pisces, and SA57) in $BRIzJK$.  Objects in these fields
were selected in the $K$ band down to $K = 20.0$, where the
completeness level is $80$\%. These objects were photometered in the
same manner as was done for the clusters in our sample \citep{sta02}.
Before being used for background correction, color distributions for
the field galaxy samples are corrected for the color-magnitude
relations derived for each cluster.  We have no $g$ or $V$ data in the
SPICES field survey and so need to interpolate magnitudes for those
bands for the clusters which have $g$ or $V$ data.  As part of the
analysis of SPICES data, photometric redshifts have been determined by
A.\ Connolly using the methods of \cite{cs99} and \cite{csa00}.  This
process yields spectral types as well photometric redshifts. Using
these spectral types and the \cite{cww80} spectral energy
distributions, we may calculate colors in any band for the objects in
the SPICES catalogs.  As a test of this procedure, we show a
comparison of interpolated and measured colors in Figure~\ref{fig2}. 
The average of the differences amounts to only a few hundredths of 
a magnitude with an rms varying from 0.15 to 0.3 mags (as shown in 
the figure).  This indicates that the interpolated $g$ and $V$ band 
colors of the field galaxies in the SPICES sample are reasonably accurate. 
The above rms values are typically 2--3 times smaller than the actual 
offsets in the observed colors that are used in determining which are 
the blue galaxies when calculating Butcher-Oemler fractions.

The marginal color distributions for the clusters and the field
sample, normalized by the relative areas and with the appropriate
magnitude limits, are shown in Figure 3.  The partial galaxies in the
field histograms are due to the normalization to the area and
magnitude limit of each cluster. As we state above, we have checked
that the mode of the color distributions are close to 0 as
expected. This does not always work perfectly both because measurement
scatter in the colors tends to place fainter objects towards the
red, and because field galaxies are in the color-magnitude diagrams
biasing the fits to the cluster color-mag sequences.  The
distributions for M0906 are a special case: note they have modes
which are significantly different from 0.  This
appears to be due to the fact that this cluster is composed of two
clumps in redshift space, each with its own color-magnitude relation
\citep{ell01}.  

The arrow in each panel of Figure 3 indicates the color
corresponding to $(B-V)_{rest}= $0.2 mag bluer than the early-type
c--m relation. The blue fraction is defined as the ratio $N_b/N$,
where $N_b$ is the field-corrected number of galaxies bluer than the
BO84 color limit, and $N$ is the total number of cluster members (in a
statistical sense, after subtraction of the field population).  The
error in this quantity is derived from

$${\sigma^2 (f_B) \over f_B^2} = \Big({\sigma^2 N_b \over N^2_b}\Big)+
\Big({\sigma^2 N \over N^2}\Big)$$

and

$$N = N{\rm (all)} - N{\rm (field)}$$
$$N_b = N_b {\rm (all)} - N{\rm (field)}$$

\noindent where $N{\rm (all)}$ and $N_{\rm(field)}$ are Poisson variables
(including an extra contribution from clustering which is computed 
from the four separate background fields of SPICES) and $N_b{\rm(all)}$ and $N_b
{\rm(field)}$ are binomial variables. 
The correction for field contamination is statistical which means that
the corrections can be too large resulting, on some
occasions, in blue fractions that are negative. In these cases we have
indicated that the value is an upper limit in Figures~\ref{fig4} and
\ref{fig5} below.

We plot blue fractions as a function of redshift for the 0.7 Mpc radii
for both optical-infrared and optical-optical colors in
Figure~\ref{fig4}, and compare our results for the optical colors with
the previous compilations by BO84 and \cite{rs95} in
Figure~\ref{fig5}. The derived $f_B$'s for our $K$-selected samples
are tabulated in Table~\ref{tabn2}. A fit to our data shows that $f_B$
is about 10\% at all redshifts we consider (albeit with
large scatter).  

To see if there are any systematic differences between our blue
fractions and those reported by BO84, we have attempted to make direct
comparisons with BO84.  For this purpose, there are four clusters in
both our sample and that of BO84 for which we have adequate data to
make useful comparisons: Abell 1942, CL0024+1654, 3C~295, and
CL0016+16.  For these clusters, we have selected samples in both the
$K$-band and also in an optical band, $R$ or $I$, depending on the
cluster redshift.  Our samples are chosen over the same areas as in
BO84, and to the same magnitude limits actually used by BO84, to the
extent that these magnitude limits can be determined from the
literature.  The areas and magnitude limits for these comparisons are
summarized, along with the resulting blue fractions, in
Table~\ref{fbcomp}.  For 3 of the 4 clusters we find blue fractions in
our optically-selected samples similar to those calculated by BO84.
The exception is CL0016+16, the highest redshift cluster in BO84 at $z
= 0.54$, for which we determined a significantly higher value of
$f_B$.  However, the $f_B = 0.02 \pm 0.07$ found by BO84 is far below
the value of $f_B$ predicted by the Butcher-Oemler effect for the
cluster's redshift.  For all four clusters the $f_B$ that we found in
our $K$-selected sample in this comparison are lower than those found
from the optically-selected samples, even when the area and effective
magnitude limits are the same.  So these tests indicate that selecting
in the $K$-band is the most important factor determining the generally
lower values of $f_B$ that we find from our $K$-band selected samples.

\section{Discussion} \label{secdisc} 

We find that: (i) our infrared selected blue fractions are generally
lower than optically selected $f_B$ and (ii) our data show no strong
trend in $f_B$ with redshift.  However, at the redshifts $z < 0.6$
where our sample overlaps with that of Butcher \& Oemler, there is
little if any significant difference in $f_B$ as determined from our
$K$-selected samples and those of Butcher \& Omeler, given the larger
scatter in the $f_B$ found both by us and by Butcher \& Oemler.  

These points need further clarification. Our sample of clusters is 
heterogeneous and is somewhat biased to rich clusters at higher 
redshifts, for which blue fractions could be lower, as these objects 
are more likely to be more dynamically evolved. \cite{ae99} show that 
the BO effect weakens when X-ray selected clusters are added to the set 
of clusters in BO84. However, we have shown that the same degree of 
passive evolution is present in at least the early-type galaxies in our 
sample \citep{sed98,dep99} independent of the X-ray luminosity of the 
cluster. 

The $K$-band selection procedure adopted will identify galaxies with
colors of Sb's or later types, as long as they are above our magnitude
limit (i.e.\ are sufficiently massive).  At the redshifts we are
considering, the $R$ band passes through the rest frame $B$ and $U$
passbands, whereas $I$ samples the $V$ and $B$ bands, so our optical
colors should be sensitive even to minor episodes of star formation in
otherwise old populations.  We show a comparison of blue fractions as
derived in the optical-$K$ and optical-only colors in
Figure~\ref{fig6}.  There is no evidence of large systematic
differences in the derived blue fractions in these colors.  On average
the difference in blue fractions between optical-$K$ and purely
optical colors is $0.07 \pm 0.10$. Our $K$ selected sample is roughly
equivalent to selection by stellar mass and therefore our results
indicate that the star formation likely causing the BO effect 
takes place among relatively low mass objects as well as in normal
spirals \cite{dressler94}, \cite{oe97}.

A plausible interpretation of our results is that there are two
components to the classical BO effect.  One component, which is seen
by our $K$-selected samples giving rise to the $\sim$10\% values of
$f_B$ that we find, is the massive galaxies, with colors approximately 
equivalent to those of Sb's, that are present at a roughly constant 
fraction at all redshifts; these objects are likely to be field spirals 
whose star formation history is modified by infall into the cluster. 
The second component, which is missing from our infrared selected data, is
responsible for both the larger blue fractions of optically-selected
samples and their redshift dependence. If this is correct, these
objects have $K-$band luminosities lower than $K^* + 1.5$ and are
likely to be a population of starbursting dwarfs (or low luminosity
spirals), subject to luminosity boosting in the optical bands
\citep{bar96}. The first population would be identified with the
`normal' spirals of C94 and C98, whereas the dwarfs would constitute
the more extreme starburst and post-starburst objects.  These dwarfs
are unlikely to evolve into S0's and instead contribute to the rich
populations of dwarf galaxies observed in nearby clusters.  As dwarfs
are sampled from the more steeply rising power-law regime of the
luminosity function ($\alpha \sim -1.3$), the exact placement of an
optical magnitude limit with respect to $M^\ast$ strongly affects the
number of dwarfs included in a sample and thus can result in a large
increase in observed $f_B$. This is qualitatively consistent with the
results of C98, where the greater part of the blue population
corresponds to relatively low mass late type spirals and irregulars.

There is considerable evidence that low luminosity galaxies show 
greater star forming activity at moderate redshifts.  The Canada-France 
Redshift Survey detected a population of bright objects with strong OII 
emission at $z > 0.2$ \citep{ham97}. The blue excess field population 
at $z=0.5$ consists of dwarfs (kinematically: Mall\'en-Ornelas et al. 1997).
The luminosity function of dwarfs in two distant clusters, MS
2255.7+2039 at $z=0.29$ \citep{nfg00} and CL1601+54 at $z=0.54$
\citep{dfn01}, appears to steepen with redshift, suggesting a
brightening of the dwarf population associated with star formation
episodes. Such objects would then fade to become part of the rich dwarf
populations observed in nearby clusters \citep{wil97}. This is
consistent with the `harassment' scenario of \cite{mlk98} whereby low
mass spirals are transformed into spheroidals via interactions with the
cluster tidal field. A possible interpretation then would account for
the Butcher-Oemler effect as a cluster counterpart of the faint blue
field galaxies, coupled with an increase in cluster infall at larger
redshift, as in the scenario presented by \cite{ell01}.

Although our data only provide information on the bright, massive
component of the BO effect, we observe that there is no redshift at
which the blue fractions peak strongly in Figure~\ref{fig4}. If the
blue galaxies in our $K$-selected samples are mostly normal spirals 
falling into clusters, this suggests that there is no preferred epoch 
of cluster merging and that spiral and lenticular populations may be 
built up gradually by infall of small groups at least since $z \sim 1$. 

To conclude, we need to caution the reader that our results are based on
a heterogeneous sample of objects, which is likely to contain significant
observational biases. One obvious example is that the fit in Fig.~5 is
weighed to low blue fraction by the small number of clusters at $z > 0.7$
which have low $f_B$. If we exclude these from our fit, we still find a
shallower slope than BO84 but the discrepancy is much reduced. The higher
redshift clusters are those most likely to be massive and to have low
blue fractions because more highly evolved. This points to the the need
to acquire more data, over wider fields of view and to fainter limits,
and for a larger sample of clusters at high redshift, in order to confirm
the results presented here.
\acknowledgments 

The authors would like to thank NOAO for a generous allocation of 
observing time to this project, and the staffs at Kitt Peak and Cerro 
Tololo for their help with the observing. We wish to thank Warrick J. Couch 
for having read the manuscript and substantially improved it by 
his comments and advice. We also would like to thank the referee,
Augustus E. Oemler, for his careful and cooperative reports which
have made this paper substantially better. This research has made use 
of the NASA/IPAC Extragalactic Database (NED) which is operated by the 
Jet Propulsion Laboratory, California Institute of Technology, under 
contract with the National Aeronautics and Space Administration. 
Support for this work was provided by NASA through grant number 
AR--5790.02--94A from the Space Telescope Science Institute, which is 
operated by the Association of Universities for Research in Astronomy, 
Inc., under NASA contract NAS5-26555.   Portions of the 
research described here were carried out at the Jet Propulsion 
Laboratory, California Institute of Technology, under a contract with 
NASA.  Work performed at the Lawrence Livermore National Laboratory is 
supported by the DOE under contract W7405-ENG-48.

\clearpage 

\begin{deluxetable}{ccccc} 
\tablewidth{0pt} 
\tablecaption{Reference magnitudes vs. $z$} 
\tablehead{ 
\colhead{Redshift} & 
\colhead{$K_*$\tablenotemark{a}} & 
\colhead{$K_{lim}$\tablenotemark{b}} & 
\colhead{$K(BO)$\tablenotemark{c}} & 
\colhead{actual $K(BO)$\tablenotemark{d}} 
} 

\startdata 
0.15 & $14.84 \pm 0.49$ & 16.3 & 16.5 & 16.5 \\ 
0.20 & $15.16 \pm 0.07$ & 16.7 & 17.0 & 17.0 \\ 
0.25 & $15.64 \pm 0.38$ & 17.1 & 17.5 & 17.5 \\ 
0.32 & $15.74 \pm 0.08$ & 17.2 & 18.0 & 18.0 \\ 
0.40 & $16.50 \pm 0.11$ & 18.0 & 18.5 & 17.8 \\ 
0.46 & $16.38 \pm 0.08$ & 17.9 & 18.8 & 17.7 \\ 
0.54 & $16.85 \pm 0.18$ & 18.4 & 19.2 & 18.2 \\ 
0.69 & $17.2\tablenotemark{e}\pm 0.4$ & 18.7 & 19.1 & \nodata \\ 
0.79 & $17.51 \pm 0.26$ & 19.0 & 20.1 & \nodata \\ 
0.90 & $18.05 \pm 0.25$ & 19.5 & 20.4 & \nodata \\ 
\enddata 
\label{tab1} 
\tablenotetext{a}{Measured observed frame $K^\ast$ from De Propris et 
al.\ (1999)} 
\tablenotetext{b}{$K$ limiting magnitude used in calculating our blue 
fractions from our $K$-selected samples} 
\tablenotetext{c}{Limit in observed $K$ band corresponding to $M(V)=-20.0$} 
\tablenotetext{d}{Limit in observed $K$ corresponding to actual limit 
in $M(V)$ used by Butcher \& Oemler (1984)}
\tablenotetext{e}{Interpolated value in Figure 8 of De Propris et
al.\ (1999)}
\end{deluxetable} 
\clearpage 

\begin{deluxetable}{lccccccccc}
\tabletypesize{\footnotesize}
\rotate
\tablewidth{0pt}
\tablecaption{Comparisons with BO84}
\tablehead{
\colhead{Cluster} & 
\colhead{BO84 $f_B$}& 
\colhead{R(30)}& 
\colhead{BO84 limit\tablenotemark{a}} &
\colhead{N$_{sample}$ (BO84)\tablenotemark{b}} &
\colhead{Optical\tablenotemark{c} $f_B$} & 
\colhead{IR\tablenotemark{d} $f_B$} & 
\colhead{radius} & 
\colhead{$K_{lim}$\tablenotemark{e}} &
\colhead{N$_{field}$/N$_{sample}$} \\
\colhead{} & \colhead{} & \colhead{arcmin} & \colhead{$M(V)$}&\colhead{} & \colhead{} &
\colhead{} & \colhead{arcmin} & \colhead{}
}

\startdata
Abell 1942 & $0.17\pm 0.05$ & 2.8 & -20.0 & 57 & $0.21\pm 0.05$ & $0.16\pm 0.04$ 
& 2.2 & 17.3 & 8.7/65\\
CL0024+16  & $0.16\pm 0.02$ & 1.1 & -20.8\tablenotemark{f} & 87 & $0.16\pm 
0.02$&$0.08\pm 0.03$ & 1.1 & 17.8 & 5.7/76 \\
3C~295&$0.23\pm 0.05$&1.0&-21.1\tablenotemark{g}& 45 & $0.27\pm 0.10$ &$0.14\pm 
0.04$&1.0&17.8 & 5.5/26 \\
CL0016+16&$0.02\pm 0.07$&1.0&-21.0\tablenotemark{h}& 65 & $0.18\pm 
0.06$&$0.09\pm 0.06$&1.0&18.3 & 9/51\\
\enddata
\label{fbcomp}
\tablenotetext{a}{Actual magnitude limit used by BO84}
\tablenotetext{b}{Number of galaxies within R(30) in BO84}
\tablenotetext{c}{Blue fraction based on our optically-selected
catalogs}
\tablenotetext{d}{Blue fraction based on our $K$-band selected catalogs}
\tablenotetext{e}{$K$ limiting magnitude used in calculating our blue
fractions, which for these comparisons was adjusted so as to
reach the same limit in $M(V)$ (for a red envelope galaxy in the
cluster) as used by BO84}
\tablenotetext{f}{Butcher \& Omeler (1978) indicate that the
completness limit is $m_R = 21.8$ for their CL0024+16 data.}
\tablenotetext{g}{BO84 indicate that the actual limiting magnitude in
the rest frame $V$-band in
their study for 3C295 was $M_{lim}=-21.1$}
\tablenotetext{h}{BO84 used photometry from Koo (1981), apparently
down to $F=23.0$ which is equivalent to $M_V=-21$}
\end{deluxetable}
\clearpage

\begin{deluxetable}{lcrrrrrr}
\tablewidth{0pt}
\tabletypesize{\footnotesize}
\rotate
\tablecaption{Blue Fractions for the complete cluster sample}
\tablehead{
\colhead{Cluster ID} & \colhead{Redshift} & \colhead{$f_B (R/I-K)$} & 
\colhead{$f_B (R/I-K)$} & \colhead{$f_B$ (optical)} & \colhead{$f_B$ (optical)} & 
\colhead{$N_{field}/N_{sample}$\tablenotemark{a}} & 
\colhead{$N_{field}/N_{sample}$\tablenotemark{a}} 
\\
\colhead{} & \colhead{} & \colhead{R=0.5 Mpc} & \colhead{R=0.7 Mpc} & 
\colhead{R=0.5 Mpc} & \colhead{R=0.7 Mpc} & \colhead{R=0.5 Mpc} & 
\colhead{R=0.7 Mpc}
}
\startdata
A1146 & 0.142 & $0.000 \pm 0.000$ & $0.000 \pm 0.000$ & $-0.020 \pm 0.044$ &
$-0.024 \pm 0.046$ & 1.7/38 & 3.3/55 \\
A3305 & 0.157 & $0.050 \pm 0.056$ & $0.047 \pm 0.051$ & $0.038 \pm 0.058$ & 
$0.067 \pm 0.075$ & 4.0/24 & 7.8/29 \\
MS0906.5+1110 & 0.180 & $0.027 \pm 0.029$ & $0.018 \pm 0.019$ & $0.029 \pm 
0.045$
& $0.077 \pm 0.050$ & 5.3/42 & 10.3/65 \\
A1689 & 0.185 & $0.046 \pm 0.038$ & \nodata & $0.029 \pm 0.025$ & \nodata & 
5.2/91 & \nodata \\
A1942 & 0.224 & $0.069 \pm 0.053$ & \nodata & $0.054 \pm 0.051$ & \nodata & 
3.8/47 & \nodata \\
MS1253.9+0456 & 0.230 & $0.090 \pm 0.055$ & \nodata & $0.112 \pm 0.069$ & 
\nodata & 6.7/59 & \nodata \\
A1525 & 0.259 & $0.055 \pm 0.048$ & \nodata & $0.051 \pm 0.061$ & \nodata & 
5.7/40 & \nodata\\
M1008-1225 & 0.301 & $0.000 \pm 0.010$ & $ 0.015 \pm 0.023$ & $0.047 \pm 0.054$ &
$0.085\pm 0.066$ & 5.2/44 & 10.4/60 \\
M1147+1103 & 0.308 & $0.036 \pm 0.040$ & $0.023 \pm 0.024$ & $0.112 \pm 0.092$ &
$0.071 \pm 0.062$ & 5.2/33 & 10.3/54\\
AC118 & 0.308 & $0.073 \pm 0.054$ & $0.073 \pm 0.044$ & $0.123 \pm 0.076$ & 
$0.131
 \pm 0.062$ & 5.2/52 & 10.3/77 \\
AC114 & 0.312 & $0.050 \pm 0.051$ & $0.026 \pm 0.033$ & $0.148 \pm 0.096$ &
$0.155 \pm 0.077$ & 5.1/38 & 10.1/60 \\
AC103 & 0.313 & $0.107 \pm 0.078$ & $0.175 \pm 0.088$ & $0.082 \pm 0.064$ & 
$0.161 \pm 0.088$ & 5.1/39 & 10.0/57\\
MS2137-0234 & 0.313 & $0.014 \pm 0.057$ & $0.024 \pm 0.068$ & $0.177 \pm 0.124$ 
& $0.260 \pm 0.139$ & 5.1/27 & 10.1/36\\
Abell S0506 & 0.316 & $0.029 \pm 0.034$ & $0.069 \pm 0.047$ & $0.077 \pm 0.066$ &
$0.120 \pm 0.073$ & 5.0/38 & 9.8/52\\
M1358+6245 & 0.328 & $0.046 \pm 0.050$ & $0.084 \pm 0.056$ & $0.054 \pm 0.062$ &
$0.096 \pm 0.056$ & 4.8/39 & 9.4/59 \\
CL2244-02 & 0.330 & $-0.014 \pm 0.026$ & $-0.022 \pm 0.029$ & $0.007 \pm 0.061$ 
& $0.050 \pm 0.075$ & 4.8/26 & 9.4/36 \\
CL0024+16 & 0.391 & $0.081 \pm 0.044$ & $0.077 \pm 0.037$ & $0.153 \pm 0.068$ & 
$0.200 \pm 0.068$ & 10.2/88 & 20/119 \\
GHO 0303+1706 & 0.418 & $0.065 \pm 0.064$ & $0.086 \pm 0.059$ & $0.174 \pm 
0.096$ & $0.140 \pm 0.067$ & 9.5/46 & 18.5/74\\
3C~313 & 0.461 & $0.010 \pm 0.112$ & $0.114 \pm 0.122$ & $-0.070 \pm 0.083$ &
$0.120 \pm 0.124$ & 7.3/19 & 14.3/34\\
3C~295 & 0.461 & $0.194 \pm 0.134$ & $0.178 \pm 0.107$ & $0.240 \pm 0.152$ & 
$0.184 \pm 0.109$ & 7.3/29 & 14.4/45 \\
F1557.19TC & 0.510 & $-0.021 \pm 0.105$ & $-0.186 \pm 0.189$ & $0.143 \pm 0.132$ & 
$0.106 \pm 0.190$ & 10.6/29 & 20.7/34 \\
GHO 1601+4253 & 0.539 & $0.027 \pm 0.077$ & $0.038 \pm 0.072$ & $0.020 \pm 
0.078$ & $0.117 \pm 0.094$ & 10.1/33 & 19.8/53\\
MS0451.6-0306 & 0.539 & $0.121 \pm 0.066$ & \nodata & $0.197 \pm 0.089$ & 
\nodata
& 10.1/70 & \nodata\\
CL0016+16 & 0.545 & $ 0.000 \pm 0.043 $ & $0.049 \pm 0.051 $ & $0.193 \pm 0.098$ & 
$0.160 \pm 0.073$ & 10/52 & 19.5/80 \\
J1888.16CL & 0.560 & $0.243 \pm 0.142$ & $0.296 \pm 0.130$ & $0.353 \pm 0.181$ & 
$0.416 \pm 0.162$ & 9.7/36 & 19.1/59 \\
MS2053-0449 & 0.582 & $0.083 \pm 0.116$ & $ 0.089 \pm 0.085$ & $0.226 \pm 0.168$ &
$0.314 \pm 0.142$ & 9.4/30 & 18.4/58 \\
3C 34  & 0.689 & $-0.067 \pm 0.092$ & $0.114 \pm 0.109$ & $0.143 \pm 0.153$ & 
$0.398 \pm 0.193$ & 9.3/28 & 18.1/47 \\
GHO 1322+3027 & 0.751 & $-0.180 \pm 0.188$ & $0.018 \pm 0.152$ & $-0.200 \pm 
0.231$ & $0.089 \pm 0.173$ & 12.9/26 & 25.2/49  \\
MS1137.5+6625 & 0.780 & $ -0.006 \pm 0.099 $ & $-0.040 \pm 0.120 $ & $0.014 \pm 
0.121$ & $-0.101 \pm 0.137$ & 12.6/38 & 24.7/55 \\
MS1054.5-0327 & 0.820 & $0.052 \pm 0.065$ & $0.105 \pm 0.057$ & $ 0.013 \pm 
0.076$ & $0.009 \pm 0.069$ & 12.2/54 & 24.1/86\\
GHO 1603+4313 & 0.895 & $-0.070 \pm 0.191$ & $-0.363 \pm 0.361$ & 
$-0.310 \pm 0.236$ & $ -0.987 \pm 0.653$ & 17.1/33 & 33.5/48  \\
GHO 1603+4329 & 0.920 & $-0.297 \pm 0.393$ & $-0.342 \pm 0.251$ & $0.050 \pm 
0.190$ & $0.014 \pm 0.144$ & 15.5/24 & 30.4/54 \\

\enddata
\label{tabn2}
\tablenotetext{a}{$K-$selected}
\end{deluxetable}

\clearpage

\begin{figure} 
\figurenum{1} 
\caption{Color--magnitude diagrams, with best fits to the
color-magnitude relation shown by the solid line, in optical-$K$
vs. $K$ and in an optical color ($B-R$, $g-$, $V-I$ or $R-I$)
straddling the 4000 \AA\ break vs. $K$ for all sample clusters.
The data are for the 0.7 Mpc regions described in the text (unless 
otherwise specified) and to the magnitude limit referred to in 
Table~\ref{tab1}.}  
\epsscale{0.9}
\plotone{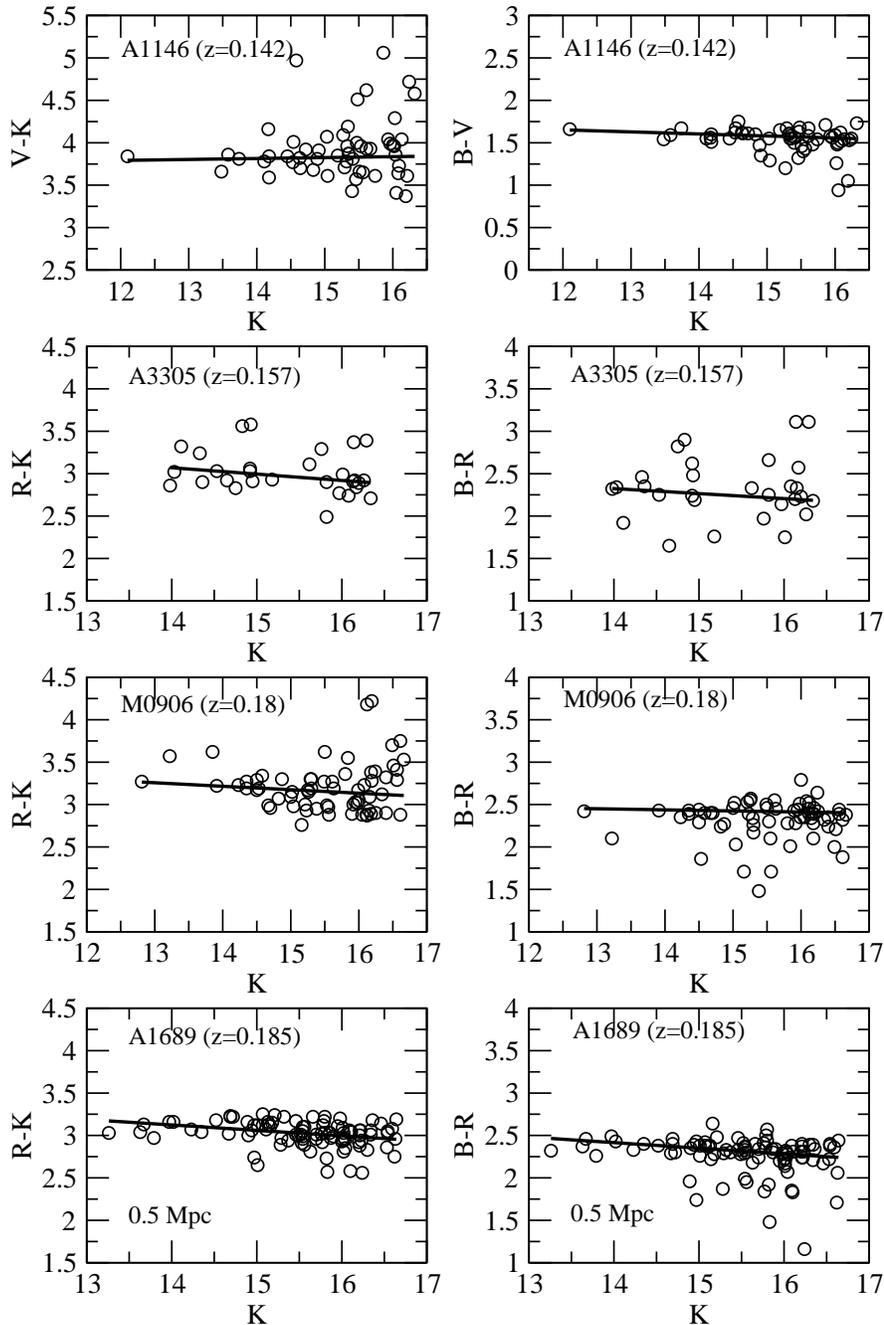}
\label{fig1} 
\end{figure} 
\clearpage 

\begin{figure} 
\figurenum{1} 
\caption{continued} 
\plotone{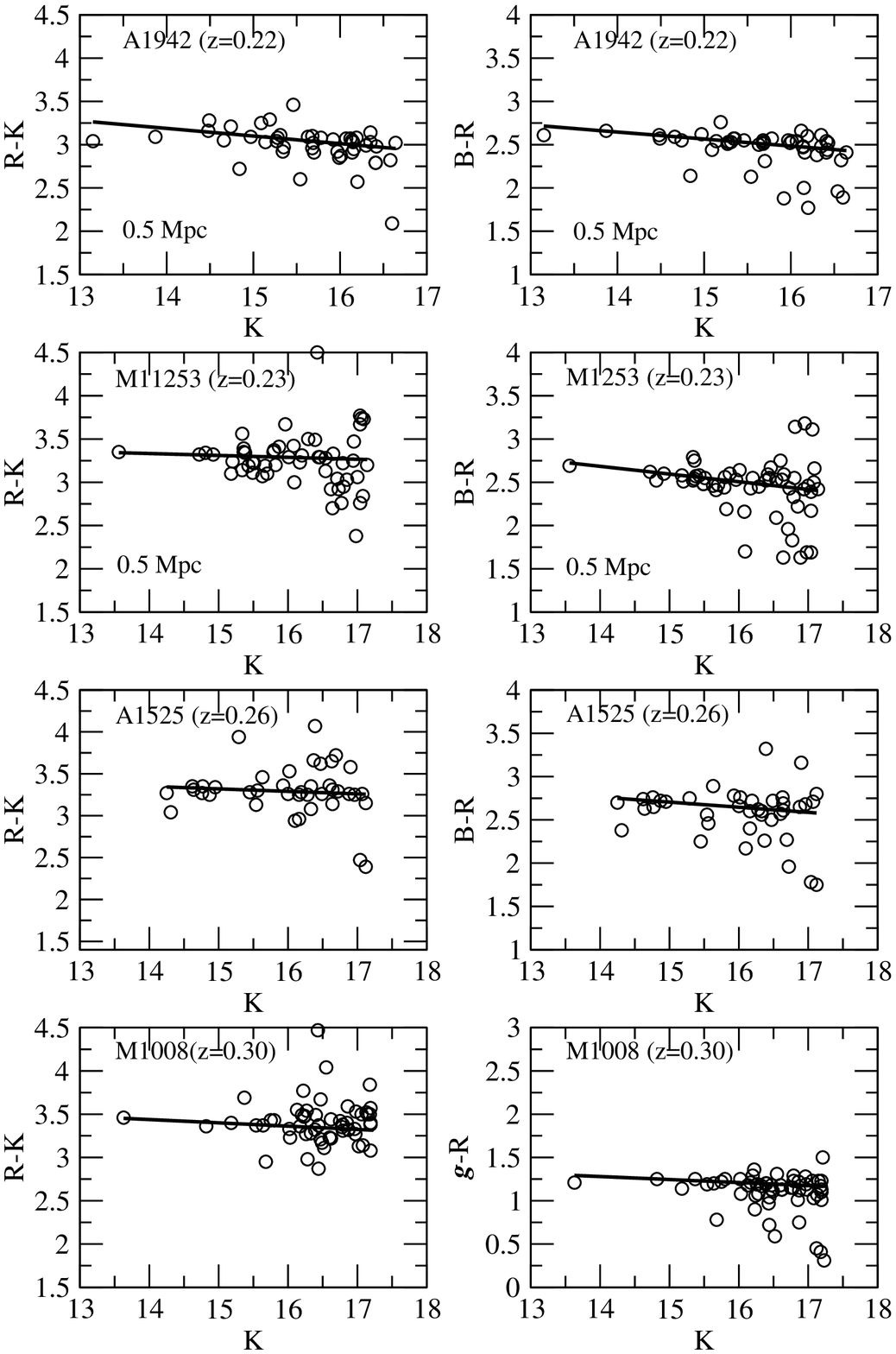} 
\end{figure} 
\clearpage 

\begin{figure} 
\figurenum{1} 
\caption{continued} 
\plotone{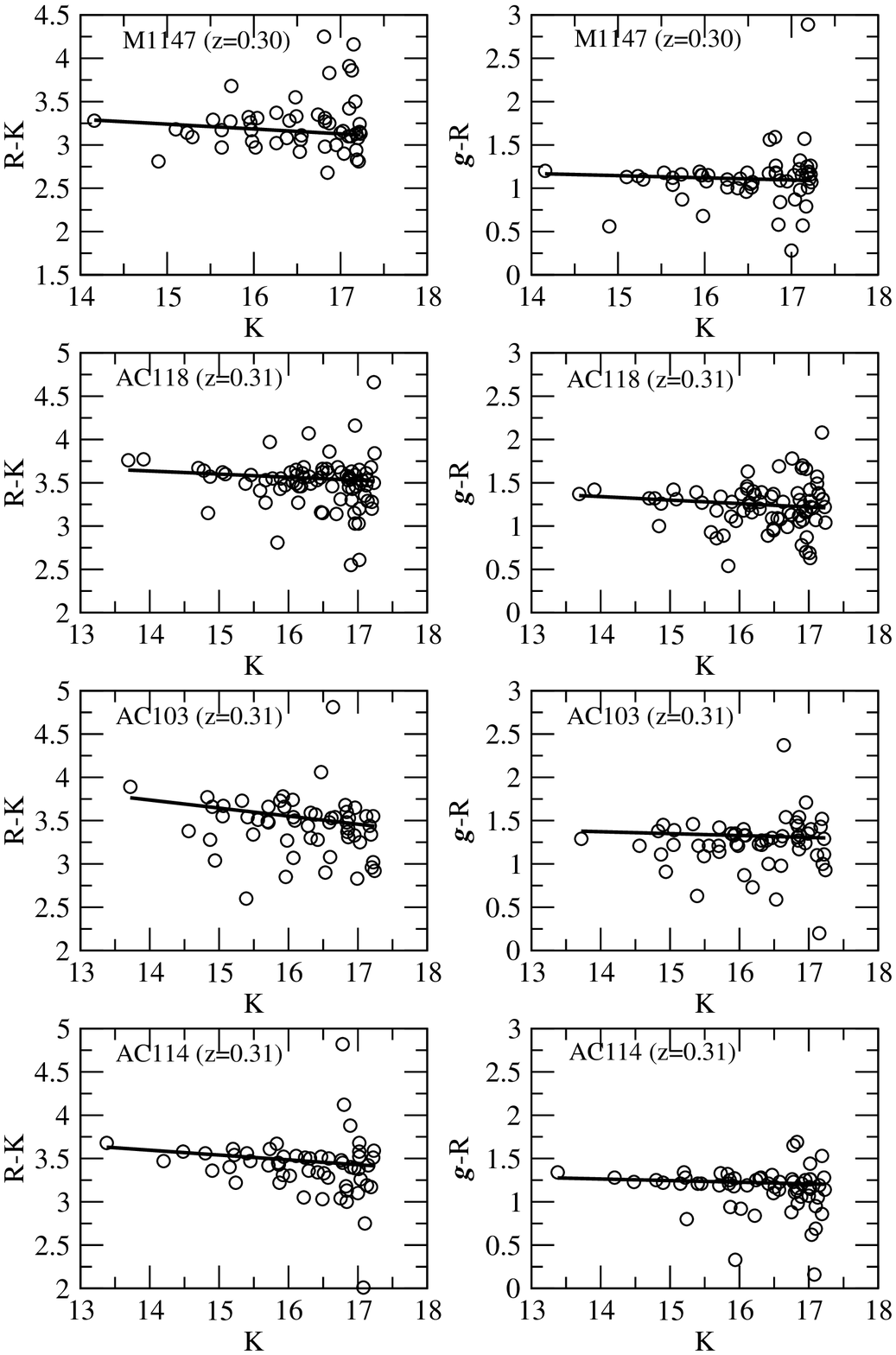} 
\end{figure} 
\clearpage 

\begin{figure} 
\figurenum{1} 
\caption{continued} 
\plotone{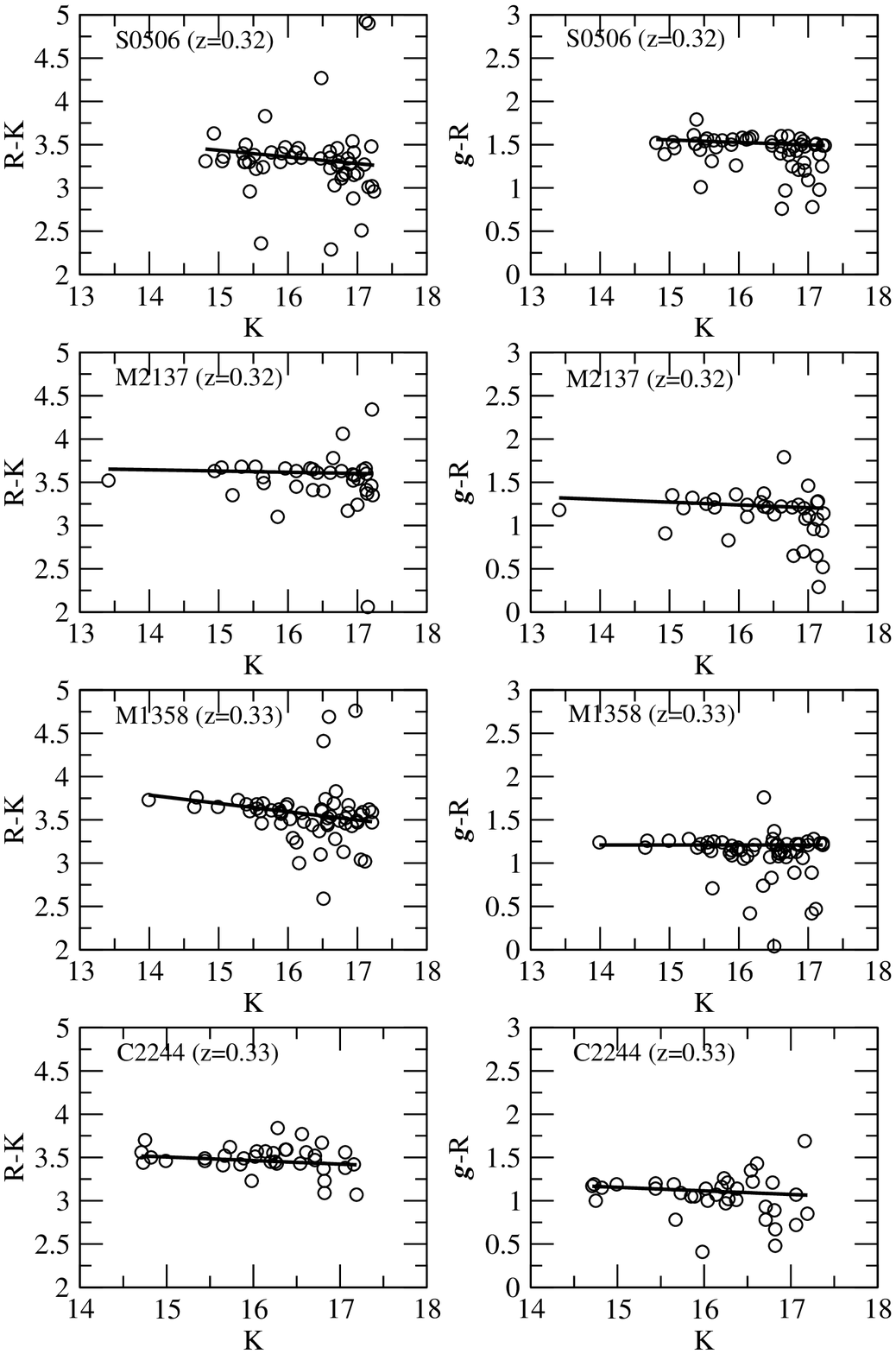} 
\end{figure} 
\clearpage 

\begin{figure} 
\figurenum{1} 
\caption{continued} 
\plotone{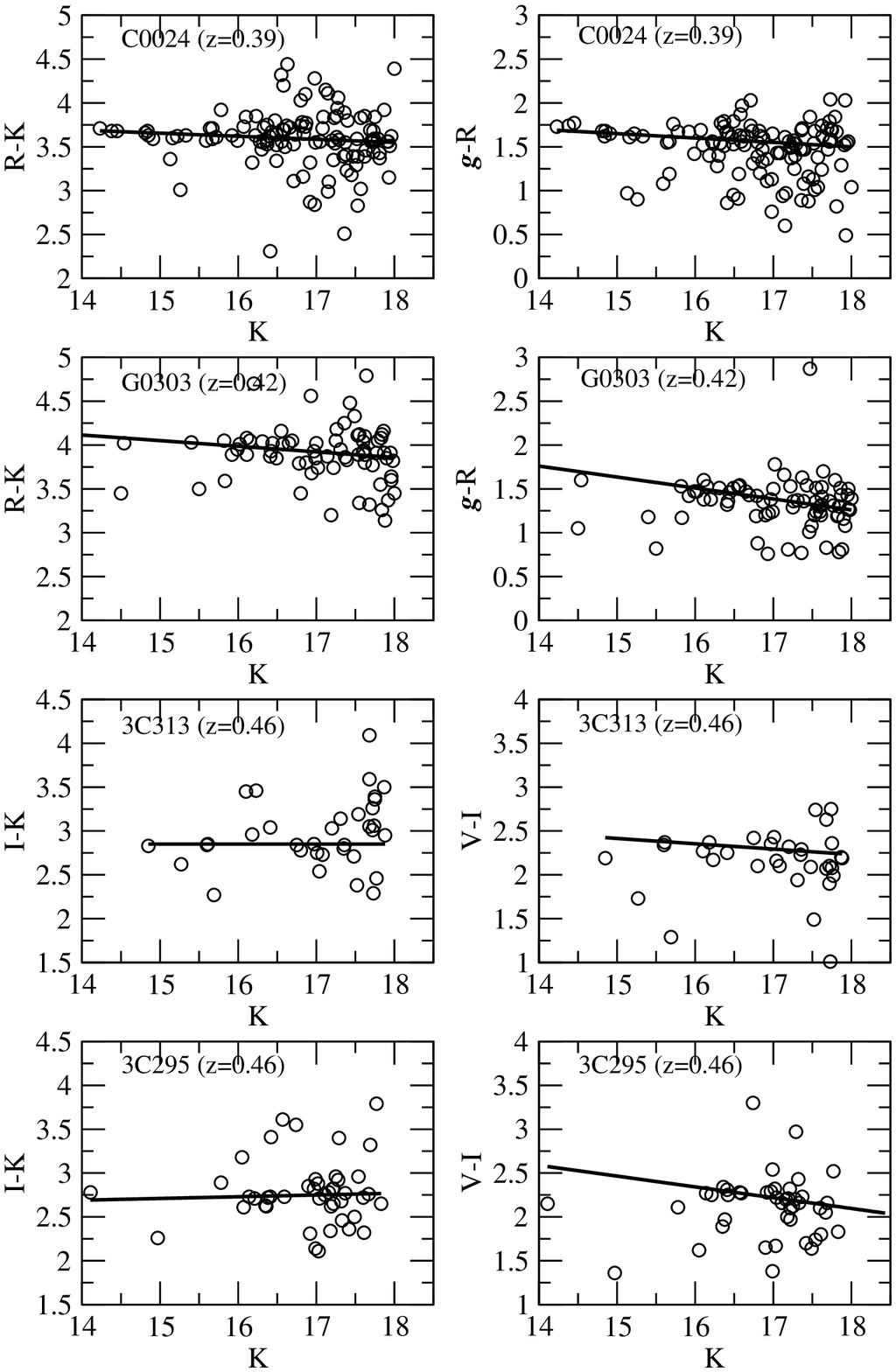} 
\end{figure} 
\clearpage 

\begin{figure} 
\figurenum{1} 
\caption{continued} 
\epsscale{0.9} 
\plotone{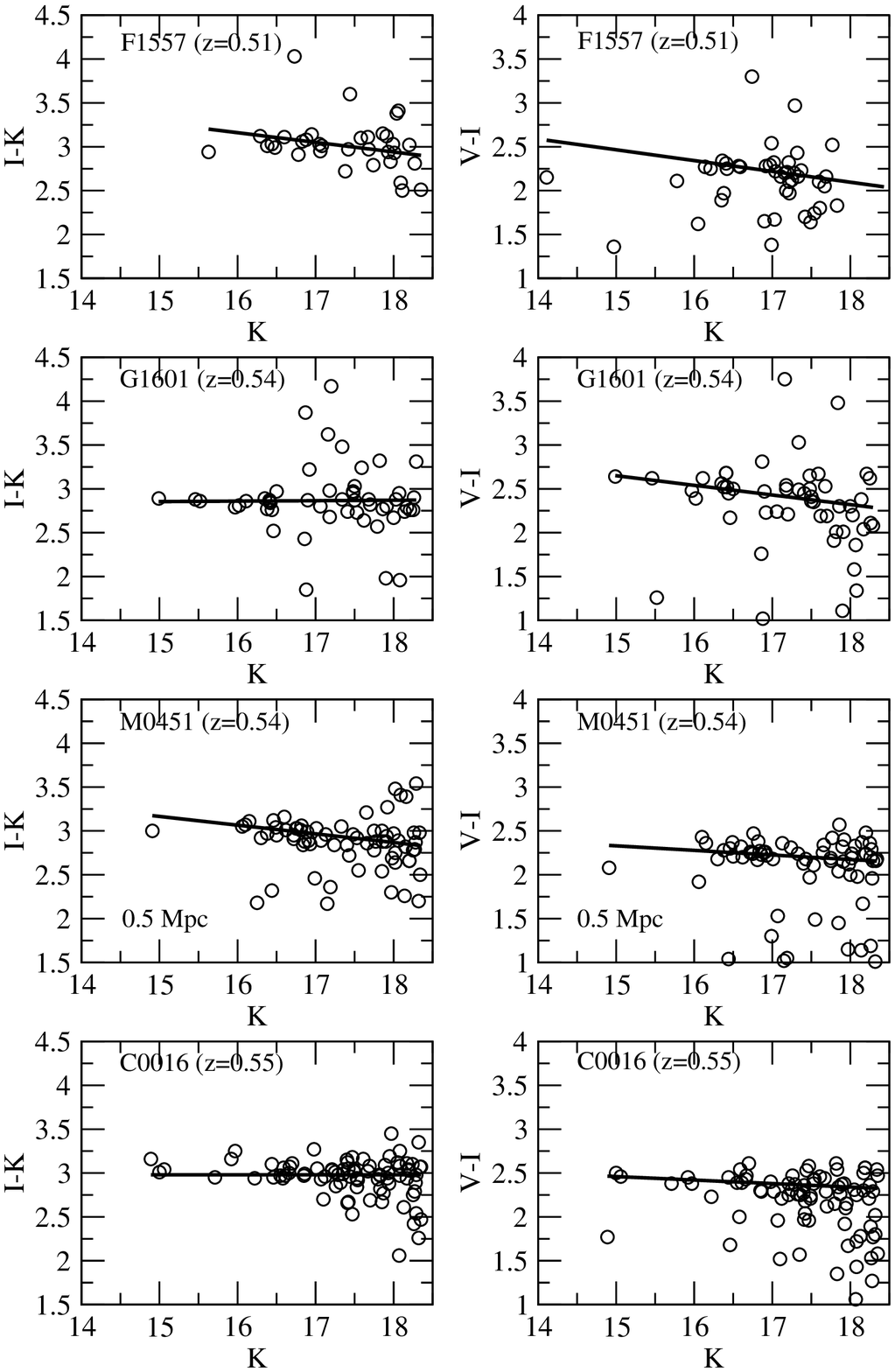}  
\end{figure} 
\clearpage 

\begin{figure} 
\figurenum{1} 
\caption{continued} 
\plotone{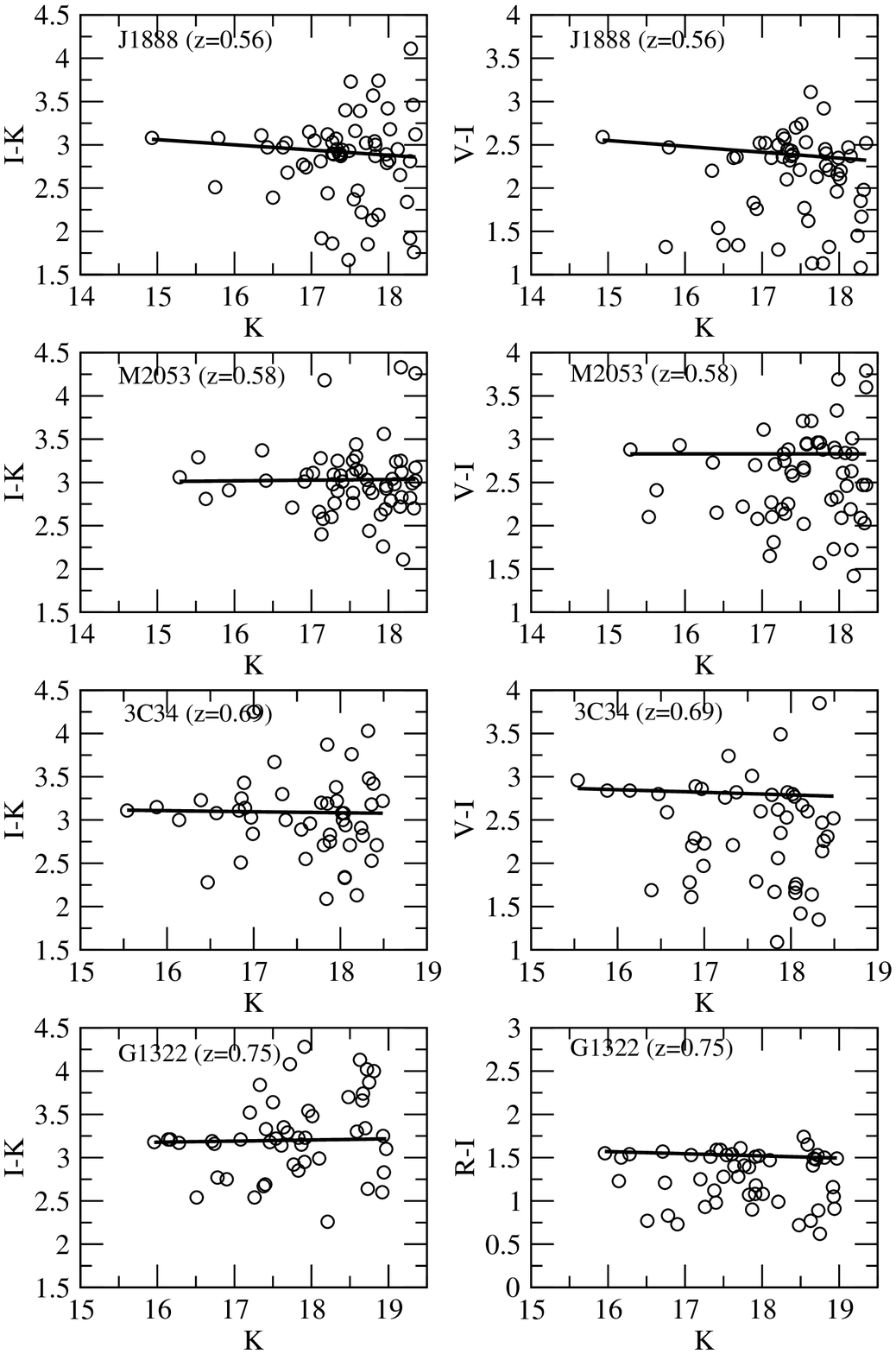} 
\end{figure} 
\clearpage 

\begin{figure} 
\figurenum{1} 
\caption{continued} 
\plotone{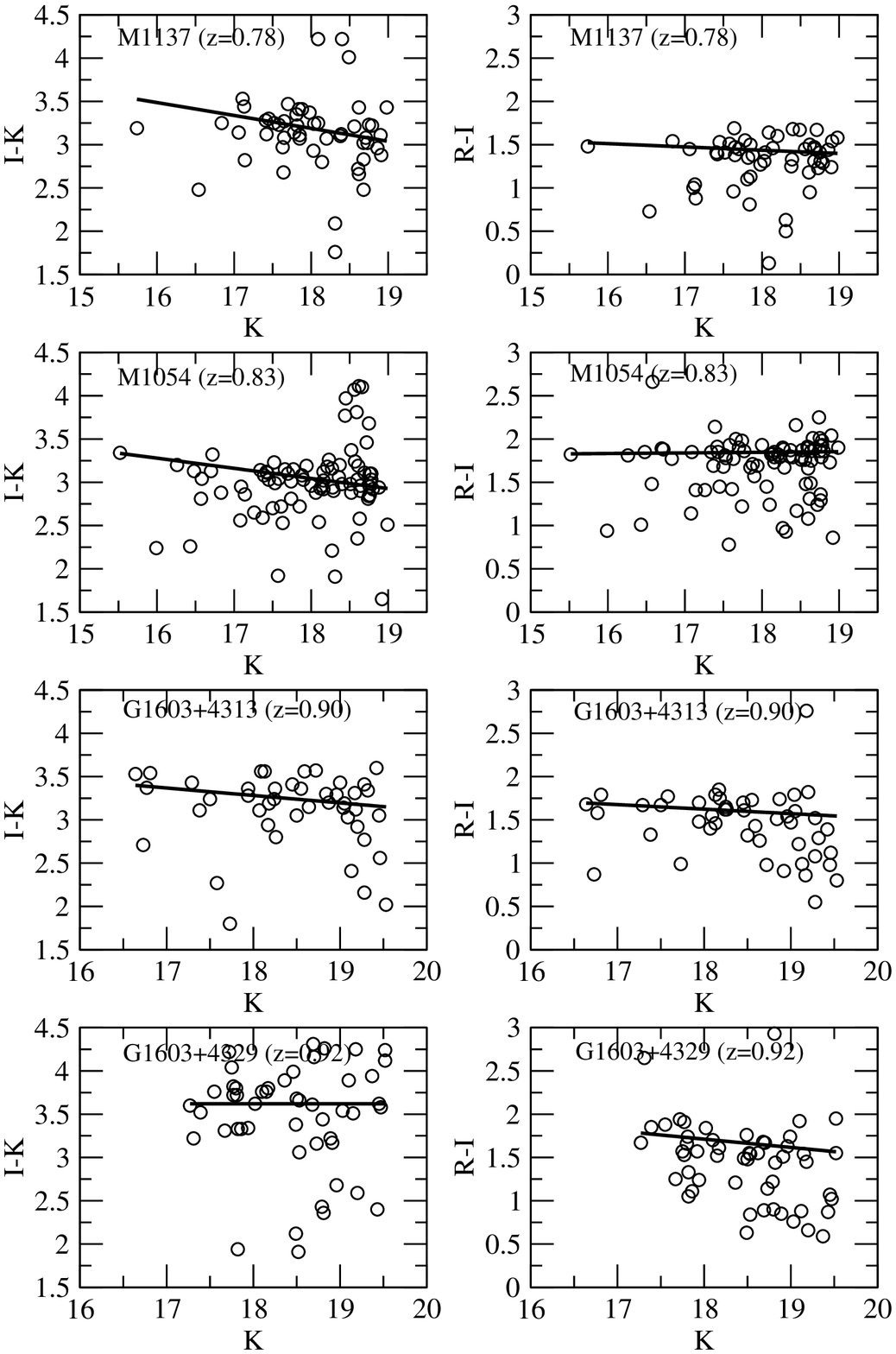} 
\end{figure} 
\clearpage 

\begin{figure} 
\figurenum{2} 
\caption{Comparison between interpolated and measured colors for 
galaxies in the SPICES survey.} 
\plotone{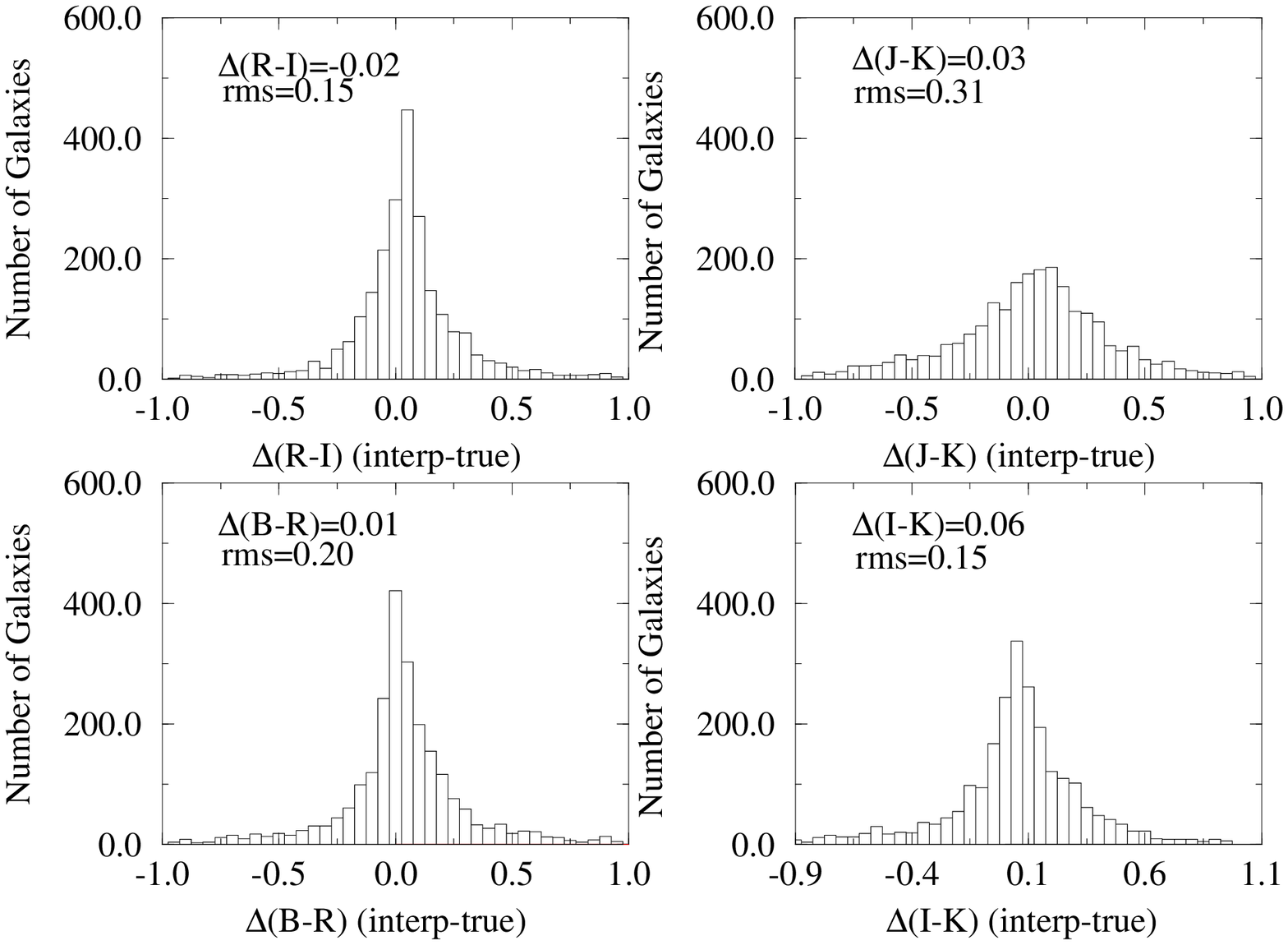} 
\label{fig2} 
\end{figure} 
\clearpage 

\begin{figure} 
\figurenum{3} 
\caption{ 
Marginal color distributions in optical-$K$ and an optical color 
straddling the 4000 \AA\  break for galaxies in the cluster fields 
(open histogram) and in the background fields (hatched histogram). We have 
removed the color--magnitude relations from both the cluster and field 
datasets and normalized field counts to the areas covered. Here we show the
distributions for the 0.7 Mpc areas we use in our analysis (except for a
few clusters where our data cover only 0.5 Mpc; these are identified in
the figure). The arrow indicates the $k$-corrected color for galaxies to 
be considered `blue' as per the definition of BO84 (see text for 
details).  A Gaussian is shown centered at 0 in the marginal color
with a width specified by the expected scatter in the observed colors
due only to measurement uncertainties. 
} 
\label{fig3} 
\plotone{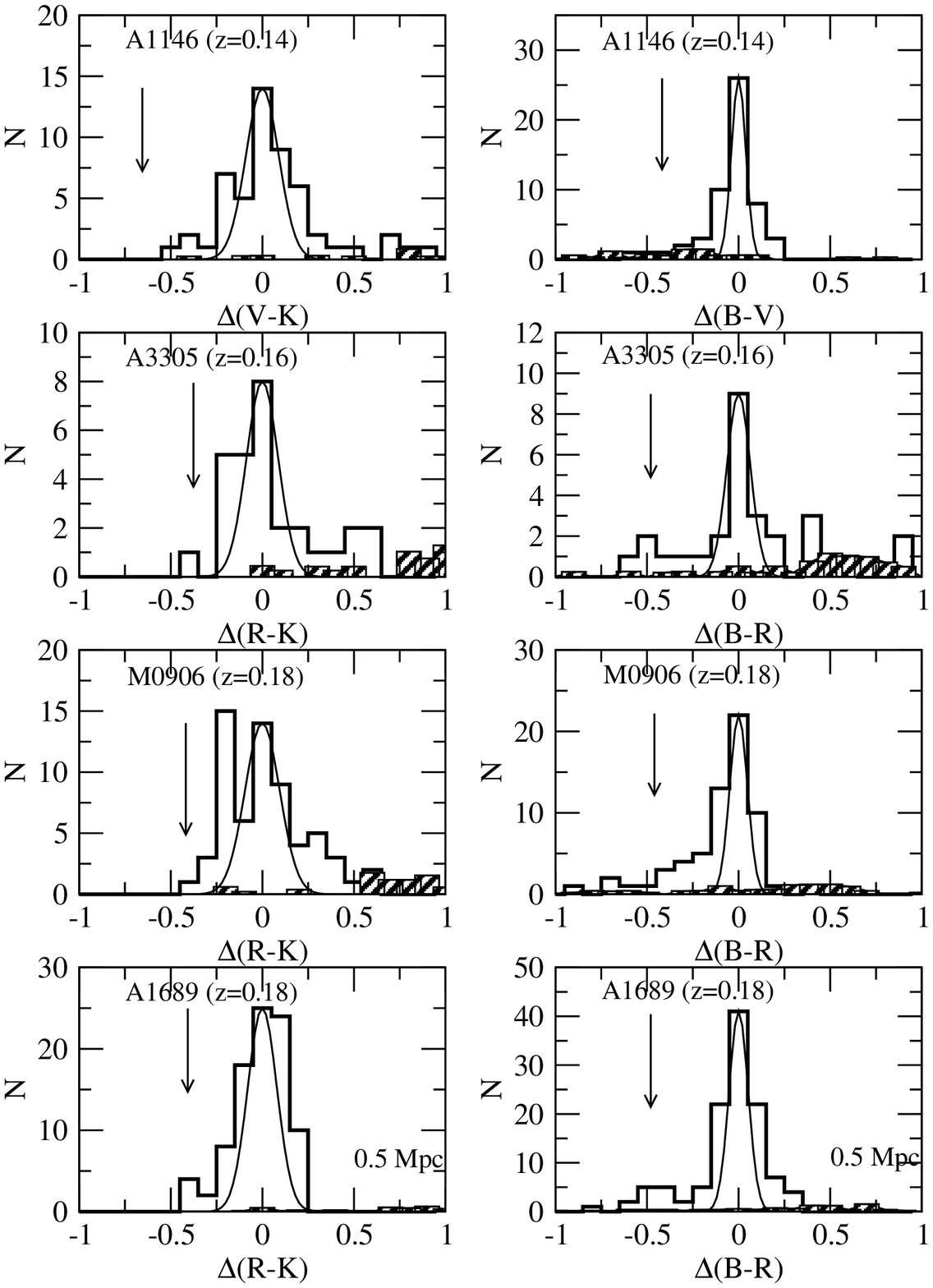} 
\end{figure} 
\clearpage 

\begin{figure} 
\figurenum{3} 
\caption{continued} 
\plotone{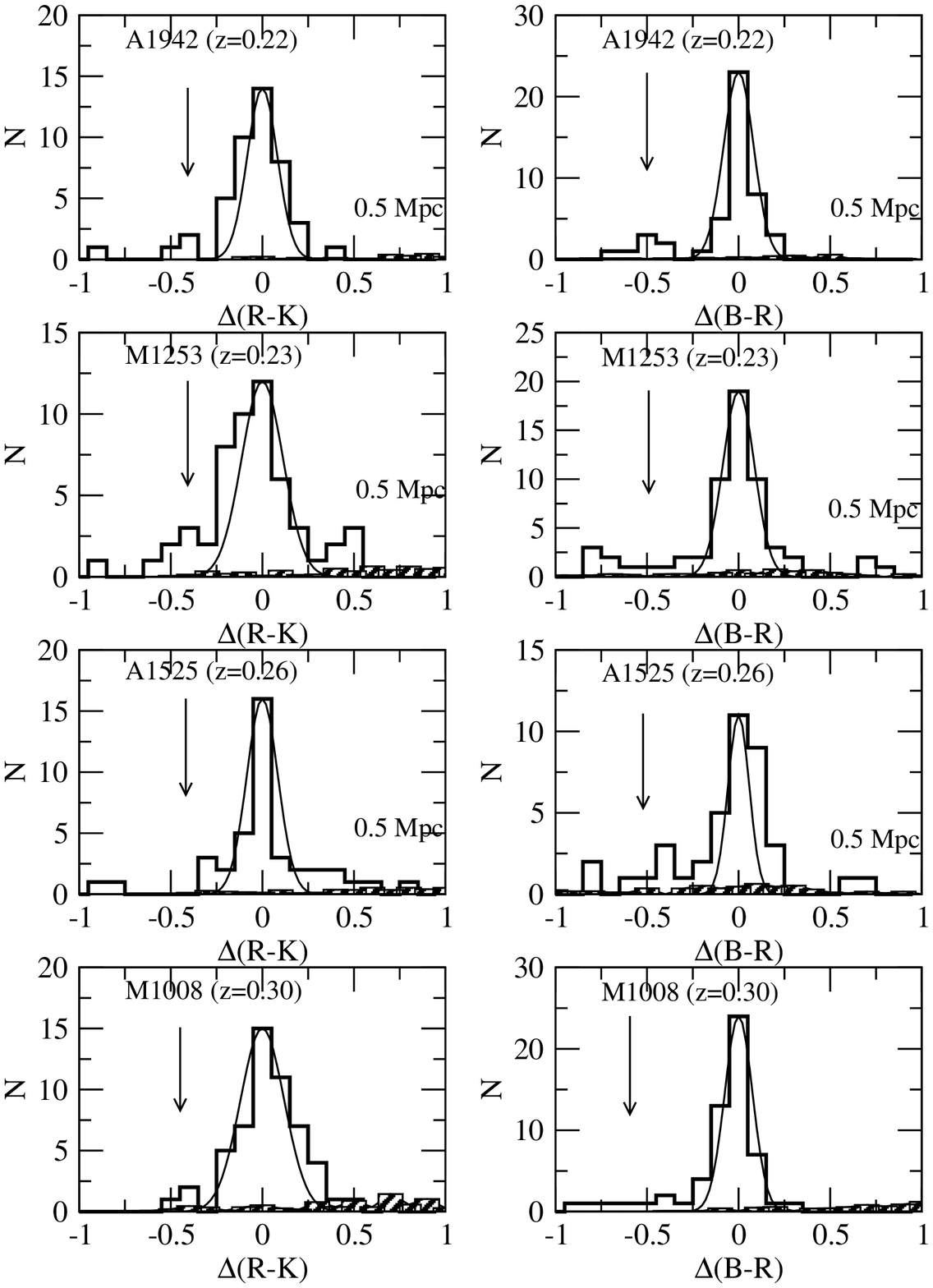} 
\end{figure} 
\clearpage 

\begin{figure} 
\figurenum{3} 
\caption{continued} 
\plotone{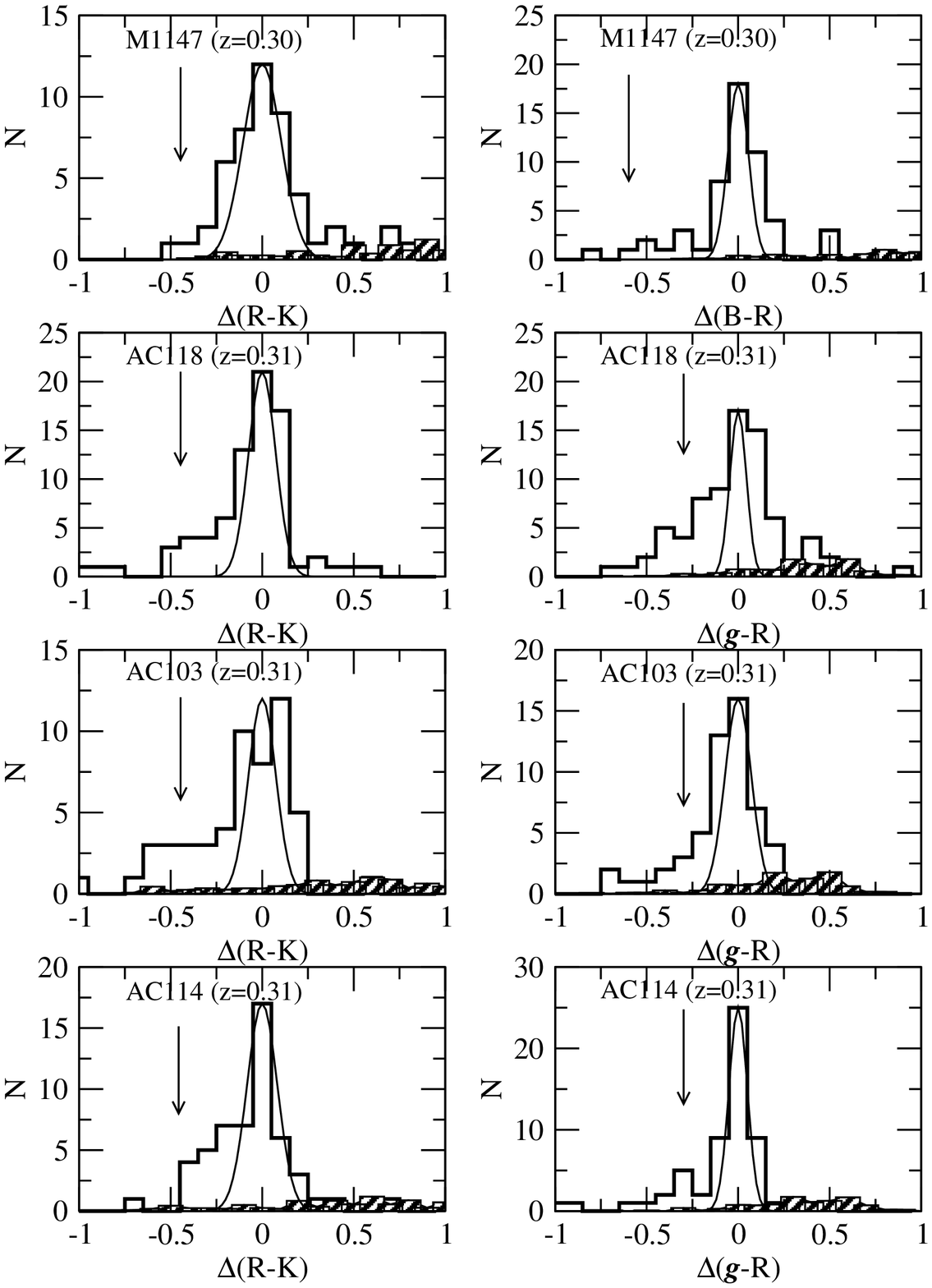} 
\end{figure} 
\clearpage 

\begin{figure} 
\figurenum{3} 
\caption{continued} 
\plotone{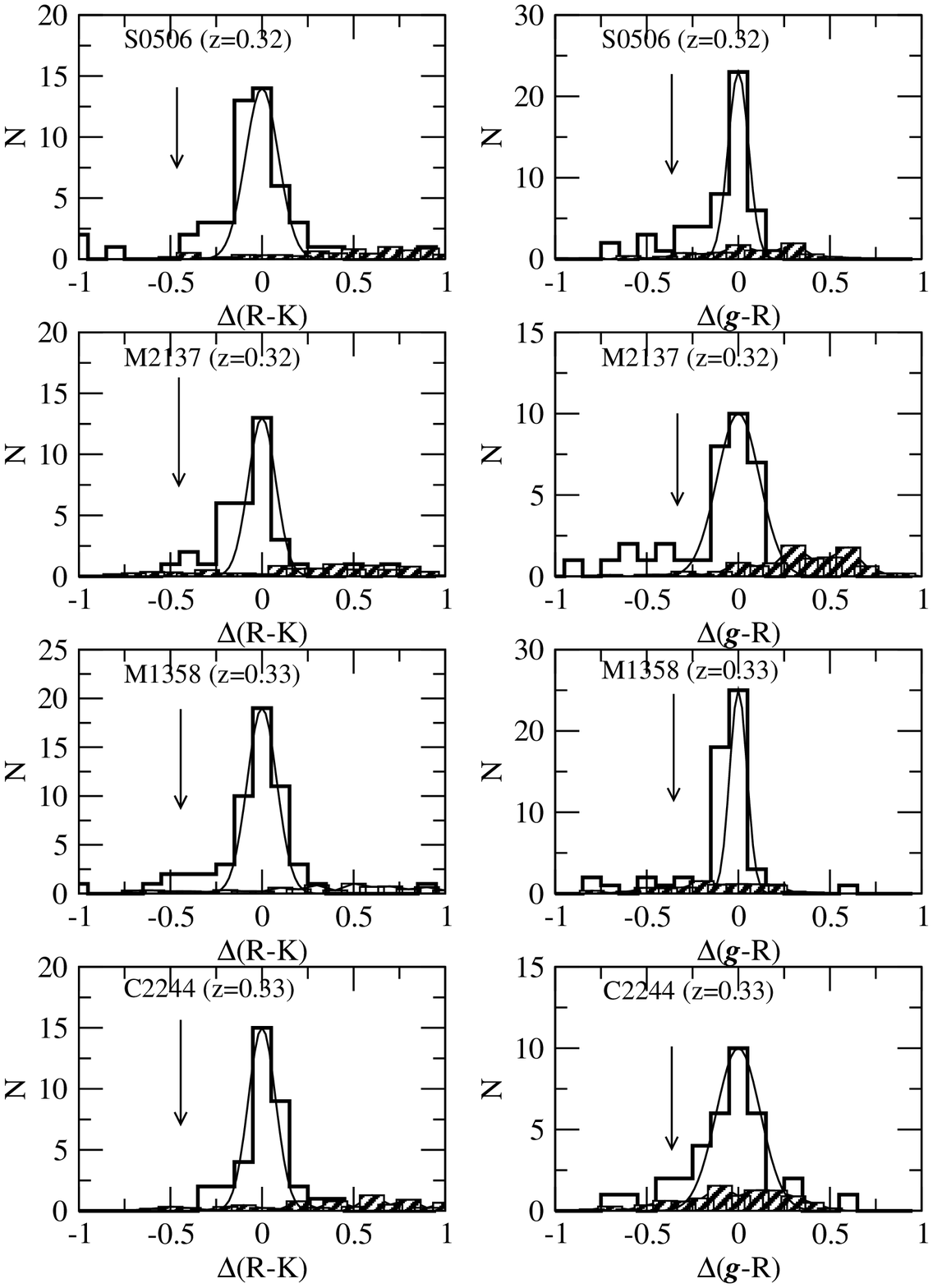} 
\end{figure} 
\clearpage 

\begin{figure} 
\figurenum{3} 
\caption{continued} 
\plotone{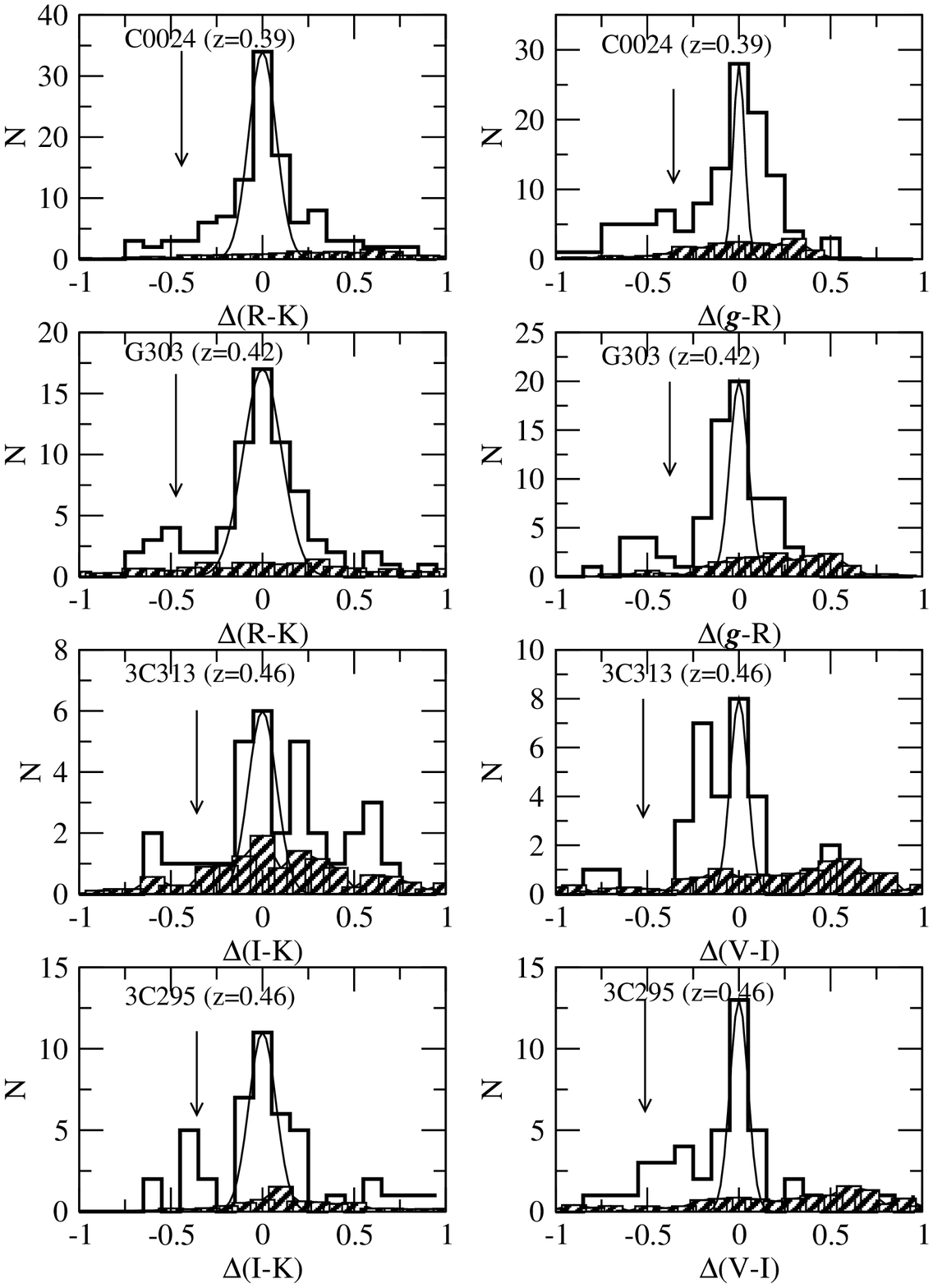} 
\end{figure} 
\clearpage 

\begin{figure} 
\figurenum{3} 
\caption{continued} 
\plotone{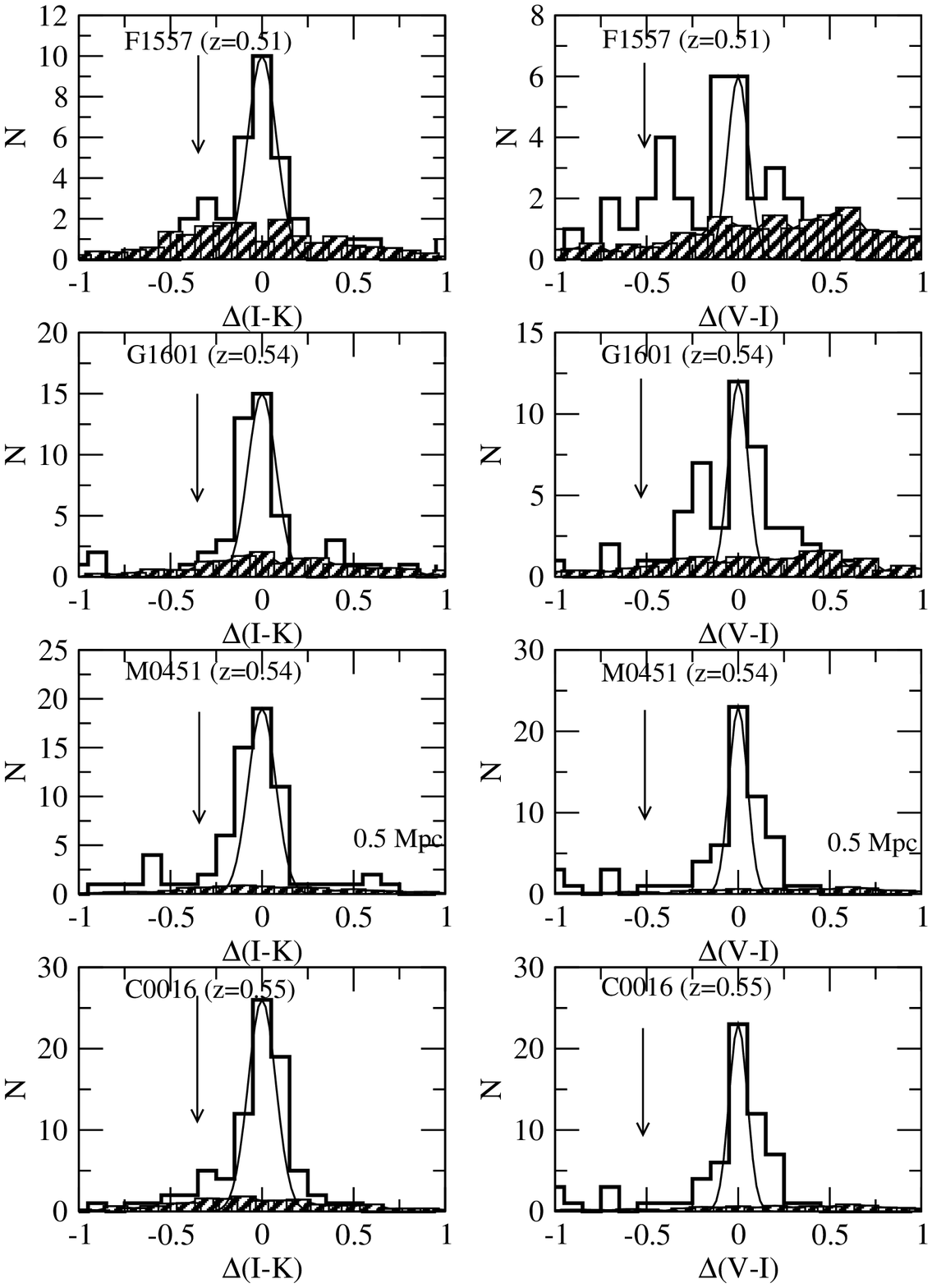} 
\end{figure} 
\clearpage 

\begin{figure} 
\figurenum{3} 
\caption{continued} 
\plotone{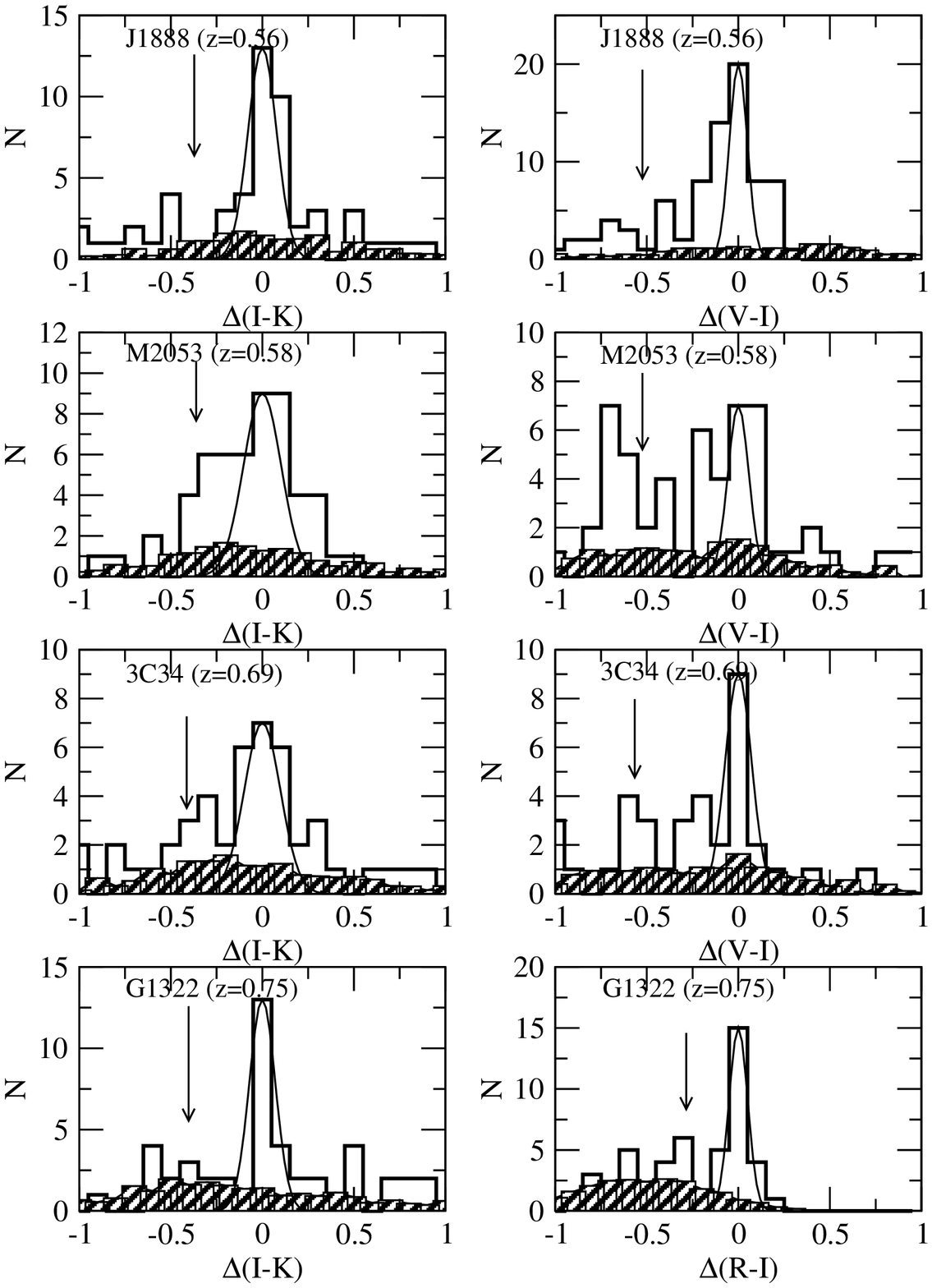} 
\end{figure} 
\clearpage 

\begin{figure} 
\figurenum{3} 
\caption{continued} 
\plotone{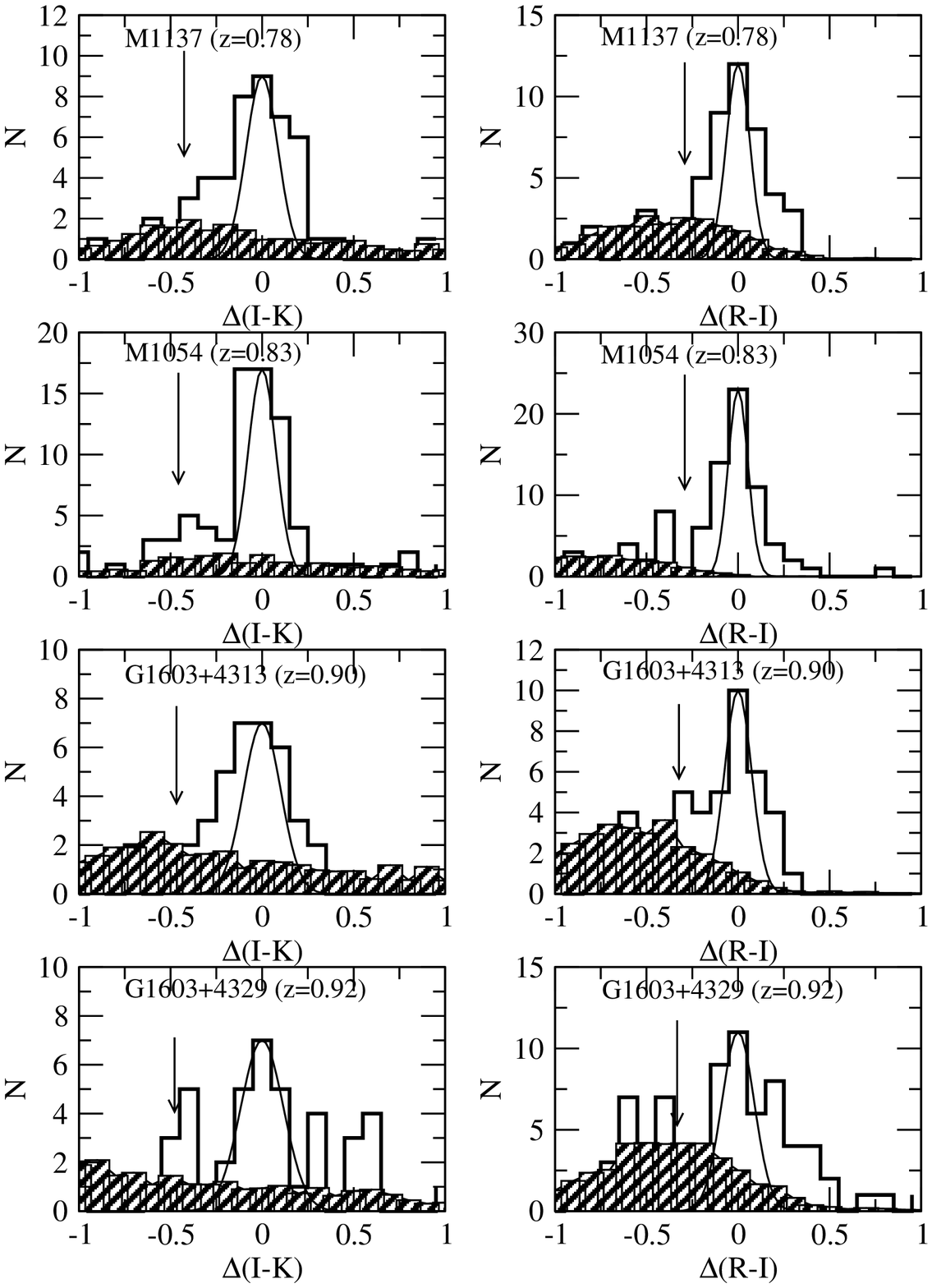} 
\end{figure} 
\clearpage 

\begin{figure} 
\figurenum{4} 
\caption{ 
Variation of the blue fraction with redshift in the 
0.7 Mpc fields in all clusters  for both the optical$-K$ 
(bottom panel) and the optical color (top panel). The upper
limits (downward pointing arrows) indicate clusters for which
the derived f$_B$ is negative because of overcorrection of the
field population.} 
\plotone{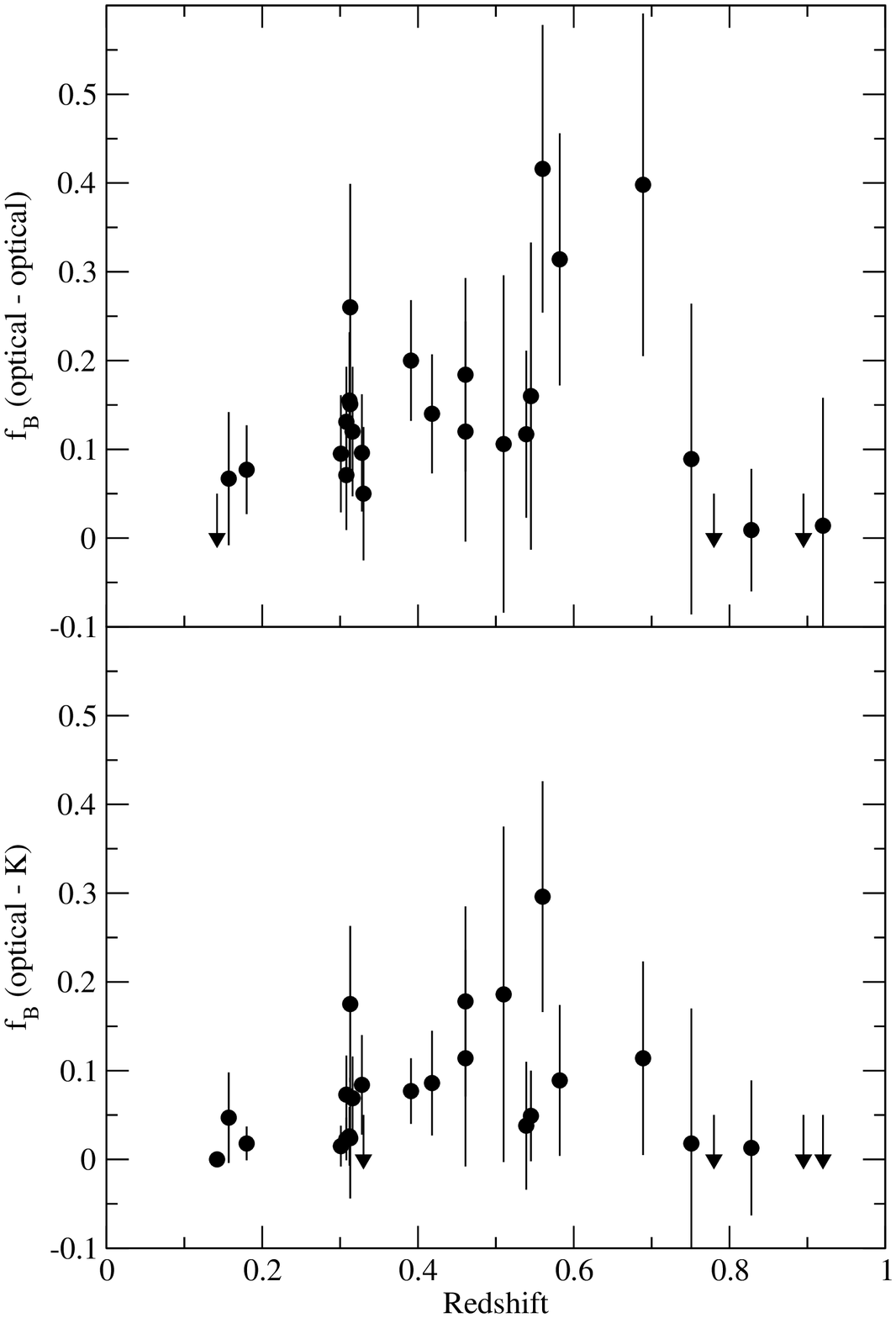} 
\label{fig4} 
\end{figure} 
\clearpage 

\begin{figure} 
\figurenum{5} 
\caption{Comparison of blue fractions in our optical colors with BO84
and \cite{rs95} for the R=0.7 Mpc fields. Closed symbols are our data;
open circles show BO84's and open squares \cite{rs95}'s data. Upper
limits as for Figure 4 above. The thick solid line is a linear fit to
our data to show the trend in blue fraction with redshift. The thin
dot dashed line is the fit shown by BO84 to their data, extrapolated
to $z=1$. }  
\plotone{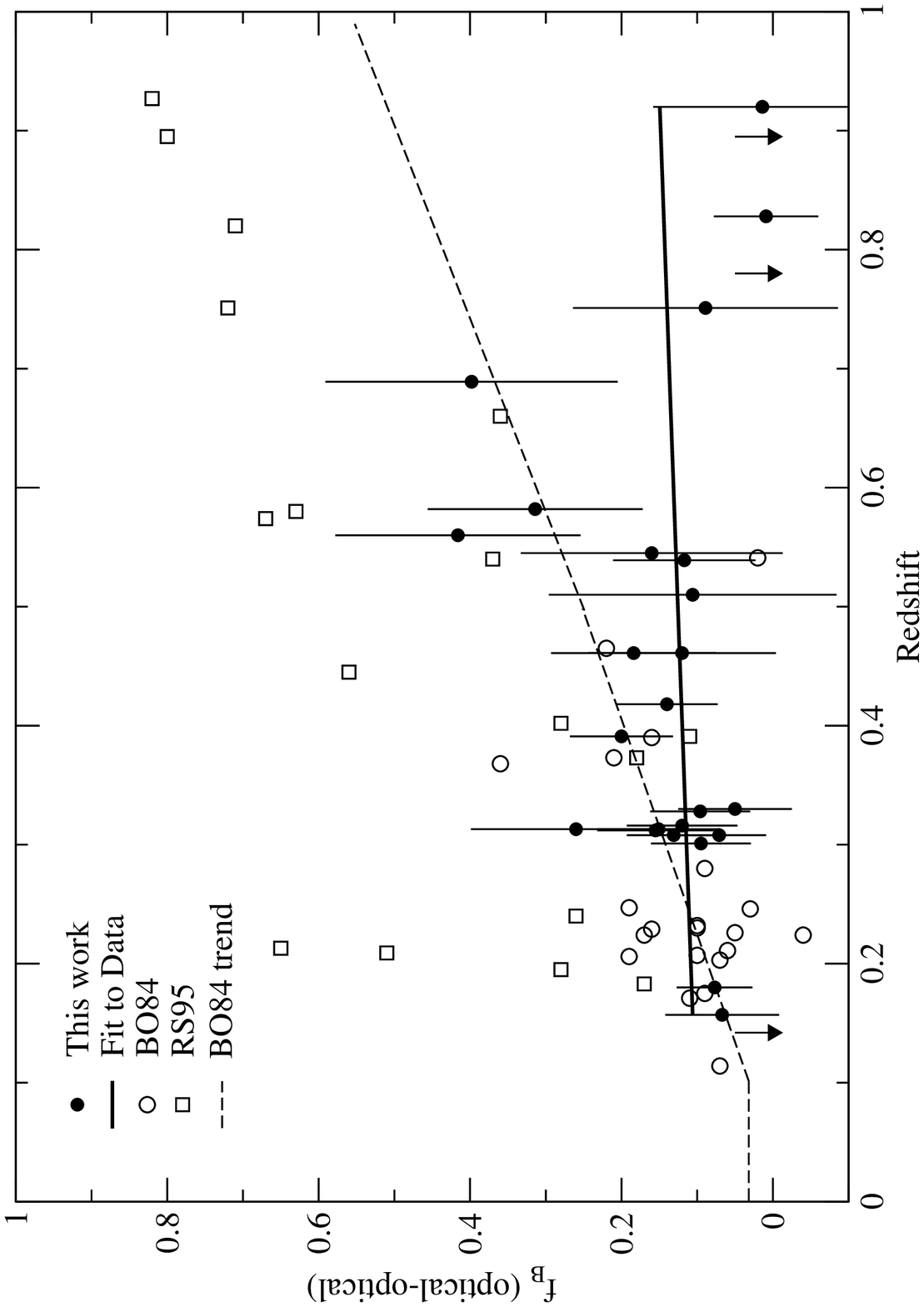}
\label{fig5} 
\end{figure} 
\clearpage 

\begin{figure} 
\figurenum{6} 
\caption{Differences in blue fractions between optical-only and 
$R/I-K$ colors as a function of redshift for the R=0.7 Mpc 
aperture. The thick solid line is a straight line fit to these 
differences.}   
\plotone{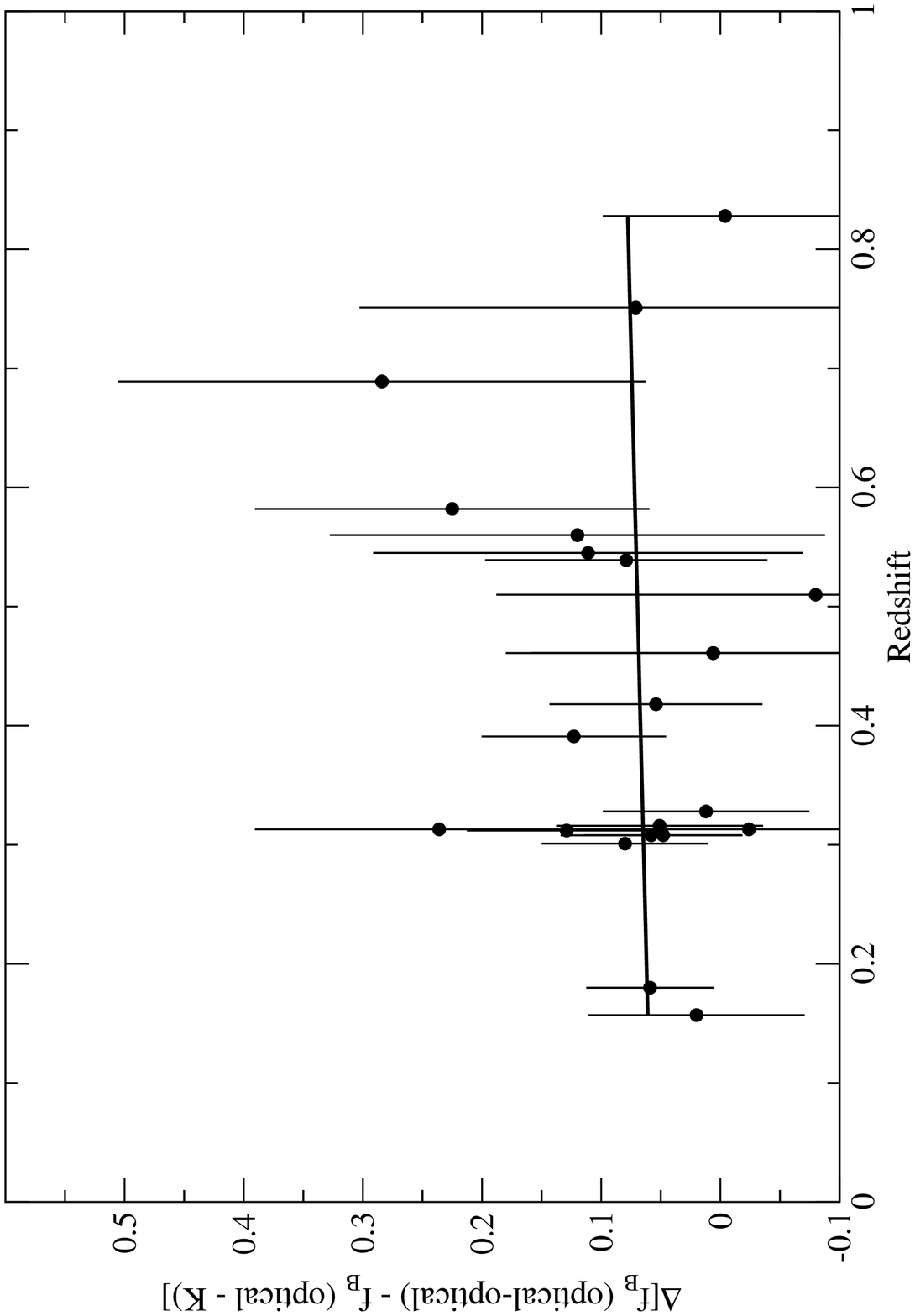} 
\label{fig6} 
\end{figure} 
\clearpage 


\begin{thebibliography}{100} 
\expandafter\ifx\csname natexlab\endcsname\relax\def\natexlab#1{#1}\fi 

\bibitem[{{Abraham} {et~al.}(1996)}]{abr96} 
{Abraham}, R. {et~al}, 1996, \apj, 471, 694 

\bibitem[{{Andreon} \& {Ettori}(1999)}]{ae99} 
{Andreon}, S., \& {Ettori}, S. 1999, \apj, 516, 647 

\bibitem[{{Aragon-Salamanca} {et~al.}(1991){Aragon-Salamanca}, {Ellis}, 
\& {Sharples}}]{aes91} 
{Aragon-Salamanca}, A., {Ellis}, R.~S. \& {Sharples}, R.~M. 1991, \mnras, 
248, 128 

\bibitem[{{Balogh} {et~al.}(2000){Balogh}, {Navarro}, \& {Morris}}]{bal99} 
{Balogh}, M. L., {Navarro}, J.~F., \& {Morris}, S.~L. 1999, \apj, 527, 54 

\bibitem[{{Barger} {et~al.}(1996){Barger}, {Aragon-Salamanca}, {Ellis}, 
{Couch}, {Smail}, \& {Sharples}}]{bar96} 
{Barger}, A.~J., {Aragon-Salamanca}, A., {Ellis}, R.~S., {Couch}, W.~J., 
{Smail}, I. \& {Sharples}, R.~M. 1996, \mnras, 279, 1 

\bibitem[{{Butcher} \& {Oemler}(1978)}]{bo78} 
{Butcher}, H., \& {Oemler}, A. 1978, \apj, 226, 559 

\bibitem[{{Butcher} \& {Oemler}(1984)}]{bo84} 
{Butcher}, H., \& {Oemler}, A. 1984, \apj, 285, 426 

\bibitem[{{Coleman} {et~al.}(1980){Coleman}, {Wu}, \& {Weedman}}]{cww80} 
{Coleman}, G.~D., {Wu}, C.-C., \& {Weedman}, D.~W. 1980, \apjs, 43, 393 

\bibitem[{{Connolly} \& {Szalay}(1999)}]{cs99} 
{Connolly}, A.~J. \& {Szalay}, A.~S. 1999, \aj, 117, 2052 

\bibitem[{{Couch} \& {Sharples}(1987)}]{cs87} 
{Couch}, W.~J., \& {Sharples}, R.~M. 1987, \mnras, 229, 483 

\bibitem[{{Couch} {et~al.}(1994){Couch}, {Ellis}, {Sharples}, \& 
{Smail}}]{co94} 
{Couch}, W.~J., {Ellis}, R.~S., {Sharples}, R.~M. \& {Smail}, I. 1994, \apj, 
430, 121 

\bibitem[{{Couch} {et~al.}(1998){Couch}, {Barger}, {Smail}, {Ellis}, \& 
{Sharples}}]{co98} 
{Couch}, W.~J., {Barger}, A.~J., {Smail}, I., {Ellis}, R.~S. \& {Sharples}, 
R.~M. 1998, \apj, 497, 188 

\bibitem[{{Csabai} {et~al.}(2000){Csabai}, {Connolly}, {Szalay}, \& 
{Bud\'avari}}]{csa00} 
{Csabai}, I., {Connolly}, A.~J., {Szalay}, A.~S. \& {Bud\'avari}, T. 2000, 
\aj, 119, 69 
  
\bibitem[{{D\'ahlen} {et~al.}(2001){D\'ahlen}, {Fransson}, \& 
{N\"aslund}}]{dfn01} 
{D\'ahlen}, T., {Fransson}, C. \& {N\"aslund}, M. 2001, \mnras, 330, 167 

\bibitem[{{De~Propris} {et~al.}(1999){De~Propris}, {Stanford}, {Eisenhardt}, 
{Dickinson}, \& {Elston}}]{dep99} 
{De~Propris}, R., {Stanford}, S.~A., {Eisenhardt}, P.~R., {Dickinson}, M., \& 
{Elston}, R. 1999, \aj, 118, 719 

\bibitem[{{Dressler}(1980)}]{dre80} 
{Dressler}, A. 1980, \apj, 236, 351 

\bibitem[{{Dressler} \& {Gunn}(1983)}]{dg83} 
{Dressler}, A. \& {Gunn}, J.~E. 1983, \apj, 270, 7 


\bibitem[{{Dressler} {et~al.}(1994){Dressler}, {Oemler}, {Butcher}, \&
{Gunn}}]{dressler94}{Dressler}, A., {Oemler}, A., {Butcher}, H., \& 
{Gunn}, J.E.\ 1994, \apj, 430, 107

\bibitem[{{Dressler} {et~al.}(1997)}]{dre97} 
{Dressler}, A. {et~al.} 1997, \apj, 490, 577 

\bibitem[{{Eisenhardt} {et~al.}(2003){Eisenhardt}, {De~Propris}, {Gonzales}, 
{Stanford}, {Dickinson} \& {Wang}}]{eis02a} 
{Eisenhardt}, P. R., {De~Propris}, R., {Gonzales}, A., {Stanford}, S.~A., 
{Dickinson}, M. \& {Wang}, M., in preparation 

\bibitem[{{Eisenhardt} {et~al.}(2003){Eisenhardt}, {Elston},
{Stanford}, {Stern}, {Wu}, {Connolly} \& {Spinrad}}]{eis02b} 
{Eisenhardt}, P.~R., {Elston}, R., {Stanford}, S.~A.,
{Stern}, D., {Wu}, K.~L., {Connolly}, A.~J. \& {Spinrad}, H.
2003, in preparation 

\bibitem[{{Ellingson} {et~al.}(2001){Ellingson}, {Lin}, {Yee} \& {Carlberg}}]
{ell01}
{Ellingson} E., {Lin} H., {Yee} H. K. C. \& {Carlberg} R. 2001, \apj, 547, 609
\bibitem[{{Gavazzi} {et~al.}(1996){Gavazzi}, {Pierini}, \& {Boselli}}]{gpb96} 
{Gavazzi}, G., {Pierini}, A. \& {Boselli}, D. 1996, \aap, 312, 297 

\bibitem[{{Hammer} {et~al.}(1997)}]{ham97} 
{Hammer}, F. {et~al.} 1997, \apj, 481, 49 

\bibitem[{{Kauffmann}(1995)}]{kau95} 
{Kauffmann}, G. 1995, \mnras, 274, 161 

\bibitem[{{Koo}(1981)}]{koo81}Koo, D.\ 1981, \apj, 251, L75

\bibitem[{{Larson} {et~al.} (1980){Larson}, {Tinsley}, \& {Caldwell}}]{ltc80}
{Larson}, R.~B., Tinsley, {B.~M.} \& {Caldwell}, C.~N. 1980, ApJ, 237, 692 

\bibitem[{{Lavery} \& {Henry}(1988)}]{lh88} 
{Lavery}, R.~J., \& {Henry}, J.~P. 1988, \apj, 330, 596 

\bibitem[{{Mall\'en-Ornelas} {et~al.}(1999){Mall\'en-Ornelas}, {Lilly}, 
{Crampton} \& {Schade}}]{mao99} 
{Mall\'en-Ornelas}, G., {Lilly}, S.~J., {Crampton}, D. \& {Schade}, D. 
1999, \apj, 518, L83 

\bibitem[{{Margoniner} \& {de~Carvalho}(2000)}]{maca00} 
{Margoniner}, V.~E., \& de Carvalho, R.~R. 2000, \aj, 119, 1562 

\bibitem[{{Margoniner} {et~al.}(2001){De Carvalho}, {Gal}, \&
{Djorgovski}}]{margo01}{Margoniner}, V.E., {De Carvalho}, R.R., {Gal},
  R., \& {Djorgovski}, S.G. 2001, \apj, 548, L143 

\bibitem[{{Metevier} {et~al.}(2000){Metevier}, {Romer}, \& {Ulmer}}]{mru00} 
{Metevier}, A.~J., {Romer}, A.~K. \& {Ulmer}, M.~P. 2000, \aj, 119, 1090 

\bibitem[{{Moore} {et~al.}(1998){Moore}, {Lake}, \& {Katz}}]{mlk98} 
{Moore}, B., {Lake}, G., \& {Katz}, N. 1998, \apj, 495, 139 

\bibitem[{{Morris} {et~al}(1998){Morris}, {Hutchings}, {Carlberg}, {Yee}, {Ellingson}, {Balogh}, {Abraham}, \& {Smecker-Hane}}]{mor98} 
{Morris}, S.~L., {Hutchings} J.~B., {Carlberg}, R.~G., {Yee}, H.~K.~C., 
{Ellingson}, E., {Balogh}, M.~L., {Abraham}, R.~G., {Smecker-Hane}, T.~A. 
1998, \apj, 507, 84 

\bibitem[{{N\"aslund} {et~al.}(2000){N\"aslund}, {Fransson}, \& 
{Huldtgren}}]{nfg00} 
{N\"aslund}, M., {Fransson}, C., \& {Huldtgren}, M. 2000, \aap, 356, 435 

\bibitem[{{Oemler} {et~al.}(1997){Oemler}, {Dressler}, \& {Butcher}}]{oe97} 
{Oemler}, A., {Dressler}, A. \& {Butcher}, H. 1997, \apj, 474, 561 

\bibitem[{{Poggianti}(1997)}]{pog97} 
{Poggianti}, B.~M. 1997, \aaps, 122, 399 

\bibitem[{{Poggianti} {et~al.}(1999){Poggianti}, {Smail}, {Dressler}, {Couch}, 
{Barger}, {Butcher}, {Ellis}, \& {Oemler}}]{pog99} 
{Poggianti}, B,~M., {Smail}, I., {Dressler}, A., {Couch}, W.~J., {Barger}, 
A.~J.,{Butcher}, H., {Ellis}, R.~S. \& {Oemler}, A. 1999, \apj, 578, 516 

\bibitem[{{Rakos} \& {Schombert}(1995)}]{rs95} 
{Rakos}, K.~D., \& {Schombert}, J.~M. 1995, \apj, 439, 47 

\bibitem[{{Rakos} {et~al.}(1997){Rakos}, {Odell} \& {Schombert}}]{ros97} 
{Rakos}, K.~D., {Odell}, A.~D. \& {Schombert}, J.~M. 1997, \apj, 490, 194 

\bibitem[{{Sandage} \& {Visvanathan}(1978)}]{sv78} 
{Sandage}, A. \& {Visvanathan}, N. 1978, \apj, 223, 707 

\bibitem[{{Smail} {et~al.}(1998){Smail}, {Edge}, {Ellis}, \& 
{Blandford}}]{sma98} 
{Smail}, I., {Edge}, A.~C., {Ellis}, R.~S. \& {Blandford}, R.~D. 1998, \mnras, 
293, 124 

\bibitem[{{Stanford} {et~al.}(1998){Stanford}, {Eisenhardt}, \& 
{Dickinson}}]{sed98} 
{Stanford}, S.~A., {Eisenhardt}, P.~R., \& {Dickinson}, M. 1998, \apj, 492, 461 

\bibitem[{{Stanford} {et~al.}(2002){Stanford}, {Eisenhardt}, {Dickinson}, 
{Holden}, \& {De~Propris}}]{sta02} 
{Stanford}, S.~A., {Eisenhardt}, P.~R., {Dickinson}, M., {Holden}, B.~P., 
\& {De~Propris}, R. 2002, \apjs, 142, 153

\bibitem[{{Stern} {et~al.}(2002)}]{spices2}Stern, D.\ et al. 2002, 
\aj, 123, 2223

\bibitem[{{Wilson} {et~al.}(1997){Wilson}, {Smail}, {Ellis}, \& {Couch}}]{wil97} 
{Wilson}, G., {Smail}, I., {Ellis}, R.~S., \& {Couch}, W.~J. 1997, \mnras, 284, 915 

\end{thebibliography}
\end{document}